\documentclass[pra,twocolumn,floatfix,aps]{revtex4}

\usepackage{gensymb}
\usepackage{amsmath}
\usepackage{makeidx}
\usepackage{amssymb}
\usepackage{amsmath}
\usepackage{graphicx}
\usepackage{ulem}
  
\usepackage[usenames]{color}
\usepackage{hyperref}
\usepackage[T1]{fontenc} 
\begin{document}

\title{{Dipole-mode}  and scissors-mode   oscillations of   a dipolar supersolid}

\author{Luis E. Young-S.$^1$\footnote{lyoung@unicartagena.edu.co}}

 \author{S. K. Adhikari$^2$\footnote{sk.adhikari@unesp.br,  professores.ift.unesp.br/sk.adhikari/}}

 \affiliation{$^1$Grupo de Modelado Computacional, Facultad de Ciencias Exactas y Naturales, Universidad de Cartagena, 
130014 Cartagena, Bolivar, Colombia}

\affiliation{$^2$Instituto de F\'{\i}sica Te\'orica, UNESP - Universidade Estadual Paulista, 01.140-070 S\~ao Paulo, S\~ao Paulo, Brazil}

\begin{abstract}

We study {  dipole-mode   and  scissors-mode} oscillations  of a  
harmonically-trapped  dipolar supersolid, composed of dipolar droplets arranged on a  one-dimensional (1D) 
or a two-dimensional (2D) lattice,
to establish  the robustness of its crystalline structure under translation and rotation, 
using a beyond-mean-field model including a  Lee-Huang-Yang
 interaction. The dipolar atoms are polarized along the $z$ direction with the supersolid crystalline structure lying in the $x$-$y$ plane.
   A stable {dipole-mode} oscillation was possible in case of  both  quasi-1D and quasi-2D dipolar supersolids, whereas a sustained  
    angular scissors-mode   oscillation was possible only in the case of a quasi-1D dipolar 
    supersolid between a maximum and a minimum of trap anisotropy in the $x$-$y$ plane. In both cases there was no visible deformation of the crystalline structure of the dipolar supersolid during the oscillation. 
   The theoretical estimate of the scissors-mode-oscillation frequency was in good agreement with the present results and the agreement improved  with an increase of  the number of droplets in the supersolid  and also with an increase in the confining trap frequencies.   The results of this study can be tested 
   experimentally with present knowhow.

\end{abstract}


\maketitle

\section{Introduction}
{
A supersolid \cite{sprsld,sprsld1,sprsld2,sprsld3,sprsld4,sprsld5}, or a  superfluid solid,  is a quantum state of matter simultaneously possessing the properties of both a solid and a superfluid. Hence,  a supersolid has  a spatially-periodic crystalline structure
 as a solid,  breaking continuous translational invariance, and also enjoys frictionless flow like a superfluid,  breaking  continuous gauge invariance.
 The pioneering search
of  supersolidity in  ultra-cold  $^4$He \cite{4} was not successful \cite{5}. Later, there had been theoretical suggestions   for creating a supersolid  in a Bose-Einstein condensate 
 (BEC) with finite-range atomic interaction 
\cite{xxx}, or  specifically,
in a  dipolar BEC\cite{santos,7a,7b},   and also in a   spin-orbit (SO) coupled spinor BEC \cite{7c}.
The study of supersolids has recently gained new momentum among research workers in various fields, 
after the experimental  observation of supersolids in a quasi-one-dimensional  (quasi-1D) \cite{1d4,1d2,1d7,1d6}  and quasi-two-dimensional (quasi-2D) \cite{2d2}
dipolar BEC  and in a quasi-1D SO-coupled pseudo spin-1/2 spinor BEC \cite{18,18a}.

Recently, in the pursuit of a supersolid, 
a spatially-periodic  state with a one-dimensional (1D) stripe   pattern in density   was observed
in an SO-coupled pseudo spin-1/2  BEC
of $^{23}$Na  \cite{18}   and $^{87}$Rb \cite{18a} atoms. 
Later, in theoretical studies of  a quasi-2D 
    SO-coupled spin-1   \cite{so1-1,so1-3}, spin-2 \cite{sandeep} as well as   pseudo spin-1/2 \cite{so1/2},
    spinor BEC,   the formation of      stripe, 
 square- and  hexagonal-lattice   \cite{7c,18} patterns in density was demonstrated.  
 
 In a different front, in a strongly dipolar BEC,  for the number of atoms $N$  beyond a critical value, 
high-density droplets were  observed  experimentally in a trapped BEC of  $^{164}$Dy \cite{drop1,2d3} and  $^{168}$Er \cite{drop2} atoms and 
studied theoretically \cite{drop3,blakie}. In a quasi-1D trapped   BEC of  $^{164}$Dy \cite{1d1,1d4,1d5},  $^{162}$Dy \cite{1d2,1d3,1d7}, and $^{166}$Er \cite{1d4,1d5}  atoms,  with a large $N$, a {dipolar supersolid in the form of a}
spatially-periodic arrangement  of droplets in a straight line 
was observed in different experiments \cite{1d4,1d2,1d7,1d6}  
 and also   was studied theoretically \cite{1d6,1d8}.  In a quasi-2D  trapped   BEC of $^{164}$Dy atoms,  {a dipolar supersolid in the form of a} spatially-periodic arrangement  of droplets on 
a hexagonal lattice was observed experimentally \cite{2d2} and  established theoretically \cite{2d4,other1,other2,blakieprl,other3}.  
  In a trapped quasi-2D dipolar BEC,
the formation of  honeycomb lattice,  stripe, square lattice and other
periodic  patterns   in density,  were also found
\cite{other1,other3,other2,fau,luis,luis2} in theoretical studies. }

 In the framework of a mean-field model  employing the Gross-Pitaevskii (GP) equation,
a dipolar BEC collapses for a strong dipolar interaction beyond a critical value \cite{coll}, 
and 
 a Lee-Huang-Yang \cite{lhy} (LHY)  beyond-mean-field  interaction \cite{qf1,qf2}
is necessary   in theoretical studies   to stabilize a strongly dipolar
droplet against collapse \cite{santos}. As the number of atoms $N$ in a trapped  dipolar BEC is increased, so that the density of atoms reaches a critical value, due  to the dipolar attraction, the condensate shrinks to a very small size.  However, it cannot  collapse due to the  LHY interaction and a droplet is formed \cite{2d3,drop1} in the case of an appropriate mixture of contact and dipolar interactions.
 The size of the   droplet is  much smaller than the harmonic oscillator trap lengths.
Such droplets can accommodate a maximum number of atoms \cite{drop3} for a given harmonic trap frequencies so as to attain a critical density of atoms in the condensate. 
 As the number of atoms is increased further in the dipolar BEC, multiple droplets are generated and due to an  interplay between the  dipolar repulsion in the $x$-$y$ plane and the external trapping potential, a supersolid-like arrangement of droplets on a spatially-periodic lattice emerges as the minimum-energy state
 \cite{rev,rev2,rev3}.  In spite of the name droplet, the present dipolar BEC  droplets in a strong trap  are different from recently observed \cite{binary,binary2}    nondipolar binary  BEC droplets in free space. Nevertheless, in both cases, the collapse is arrested  \cite{petrov} by a  beyond-mean-field  LHY interaction \cite{lhy}.

Although there have been many theoretical investigations of statics of dipolar supersolids \cite{1d8,blakie1,rei,bxx,saito,bar,byy,bfau,abd,xcd,rei2,x1,x2}, there have hardly been any direct study of dynamics. 
Linear {dipole-mode} and angular scissors-mode oscillations of a BEC are  earmarks of superfluidity. Scissors-mode oscillation of a single droplet \cite{sci-sym} as well as a quasi-1D  dipolar supersolid \cite{sci-y} has been studied  experimentally { and theoretically 
\cite{string}}. {In Refs. \cite{sci-y,string} the authors studied the variation of scissors-mode-oscillation frequency with a variation of the strength of the dipolar interaction relative to that of the contact interaction.
 In the present study we consider an evolution of the scissors-mode oscillation with a variation of trap frequency while the trap passes from a quasi-1D to a quasi-2D type. Hence the two studies are complimentary to each other.}  
 In the case of a dipolar supersolid,   { dipole-mode and  scissors-mode}  oscillations, without any distortion {of the lattice structure of droplets},
test both the superfluidity and the rubustness of the  crystalline structure of the dipolar supersolid 
under translation and rotation
and hence confirm the supersolidity of these states.

Inspired by the experimental study of Ref. \cite{sci-y},
to test the superfluidity and rubustness of a dipolar supersolid, in this paper we study the linear {dipole-mode}  and angular scissors-mode oscillation dynamics of a harmonically-trapped dipolar supersolid.   The  {dipole-mode} oscillation of a quasi-1D or quasi-2D dipolar supersolid is studied employing real-time propagation by giving a sudden translation of the harmonic trap along the $x$ direction.
 To study the angular scissors-mode oscillation of a quasi-1D dipolar supersolid,
 with an asymmetric trapping potential in the $x$-$y$ plane,  a sudden rotation of the harmonic trap around the $z$ direction  is applied. 
 In both cases a continued steady oscillation of the dipolar supersolid was confirmed without any visible distortion of the crystalline structure, thus establishing the superfluidity and the robustness of the  crystalline structure   of the dipolar supersolid. 
 Although, the theoretical estimate of the scissors-mode oscillation frequency  $\omega_{\mbox{th}}=\sqrt{\omega_x^2+\omega_y^2}$ \cite{scith}, where $\omega_x$ and $\omega_y$ are the angular trap frequencies along the $x$ and $y$ directions, respectively, is a good approximation to the actual frequency of oscillation, the agreement improves as the number of droplets in the dipolar supersolid increases 
or as the confining trap becomes stronger.  
  For a sustained periodic scissors-mode oscillation, 
the trap asymmetry in  the $x$-$y$ plane should lie between an upper and lower limits.
As the  asymmetry is reduced beyond the lower limit, or increased above the upper limit,  the periodic simple-harmonic scissors-mode oscillation becomes an irregular one. 
No sustained scissors-mode oscillation was found for a quasi-2D dipolar supersolid with a hexagonal or square lattice structure. { The linear dipole-mode oscillation along the $x$ direction
is simple harmonic and takes place with the frequency of the trap $\omega_x$.}
 { While studying the scissors-mode oscillation of a quasi-1D dipolar supersolid we keep $\omega_x$ and $\omega_z$ fixed maintaining $\omega_z \gg \omega_x$ and vary $\omega_y$ such that 
 $\omega_y > \omega_x$  thus generating a trap with asymmetry in the $x$-$y$ plane as required to initiate the scissors-mode oscillation. As $\omega_y$ increases from a small value to a value larger than $\omega_z$ the trap changes from a quasi-2D type to a quasi-1D type. In this fashion we study the evolution of the scissors-mode oscillation of a quasi-1D dipolar supersolid in both types of trap; in all cases the numerical frequency of the scissors-mode oscillation was smaller than its theoretical estimate. }

In Sec. \ref{II} we consider the beyond-mean-field model including the LHY interaction.  We  also present the appropriate energy functional, a minimization of which leads to this model.   
In Sec. \ref{III} we present numerical results for  {dipole-mode} oscillation of  a quasi-1D three-droplet and quasi-2D nine-droplet dipolar supersolid of $^{164}$Dy atoms after a sudden displacement of the trap.
We also present  results
for angular scissors-mode oscillation of a   quasi-1D  three- and five-droplet dipolar  dipolar supersolid.  
{A variation of the scissors-mode frequency with $\omega_y$, as the trap evolves from a  quasi-2D to quasi-1D type, is also studied.}
A breakdown of  scissors-mode oscillation of a   nine-droplet quasi-2D square-lattice and a seven-droplet triangular-lattice
dipolar supersolid is also demonstrated.  
    Finally, in Sec. \ref{IV} we present a summary of our findings.

\section{Beyond-Mean-field model}

\label{II}

We consider a  BEC of $N$ dipolar  atoms, of mass $m$ each, polarized along the $z$ axis, 
interacting through the following 
atomic   contact and dipolar  interactions   \cite{dipbec,dip,yuka}
\begin{align}
V({\bf R})= &
\frac{\mu_0 \mu^2}{4\pi}U_{\mathrm{dd}}({\bf  R})
+\frac{4\pi \hbar^2 a}{m}\delta({\bf  r-r' }),
\label{eq.con_dipInter} \\
U_{\mathrm{dd}}({\bf R}) =& \frac{1-3\cos^2 \theta}{|{\bf  r-r'}|^3},
\end{align}
where $\mu$ is the magnetic dipole moment of each atom,  $\mu_0$ is the permeability of vacuum,  $a$ is the scattering length.  Here, $\bf r \equiv \{x,y,z\}$ and $\bf r' \equiv \{x',y',z'\}$ are the positions of the two interacting dipolar atoms  
and $\theta$ is the angle made by  $\bf R\equiv r-r'$ with the  polarization
$z$ direction. In analogy with the scattering length,  the following dipolar length $a_{\mathrm{dd}}$ 
determines the  strength of dipolar 
interaction 
\begin{align}
a_{\mathrm{dd}}=\frac{\mu_0 \mu^2 m }{ 12\pi \hbar ^2}.
 \label{eq.dl}
 \end{align}
The dimensionless ratio 
 \begin{equation}
\varepsilon_{\mathrm{dd}}\equiv \frac{a_{\mathrm{dd}}}{a}
\end{equation} 
determines
the strength of the dipolar interaction relative to  the contact interaction 
and controls many properties of a dipolar BEC.
 
In this paper we base our study on a  3D beyond-mean-field model   including the LHY interaction. 
The formation of a lattice of droplets      is described by the following  3D beyond-mean-field GP equation including the  LHY interaction \cite{dipbec,dip,2d4,blakie,yuka}
\begin{align}\label{eq.GP3d}
 \mbox i \hbar \frac{\partial \psi({\bf r},t)}{\partial t} &=\
{\Big [}  -\frac{\hbar^2}{2m}\nabla^2
+U({\bf r})
+ \frac{4\pi \hbar^2}{m}{a} N \vert \psi({\bf r},t) \vert^2 \nonumber\\
&\ +\frac{3\hbar^2}{m}a_{\mathrm{dd}}  N
\int U_{\mathrm{dd}}({\bf R})
\vert\psi({\mathbf r'},t)\vert^2 d{\mathbf r}'  
\nonumber 
 \\
& +\frac{\gamma_{\mathrm{LHY}}\hbar^2}{m}N^{3/2}
|\psi({\mathbf r},t)|^3
\Big] \psi({\bf r},t), \\
U({\bf r})&=\frac{1}{2}m(\omega_x^2x^2+\omega_y^2y^2+\omega_z ^2z^2) ,
\end{align}
where 
 $\omega_x, \omega_y, \omega_z$  are the angular frequencies along $x,y,z$ directions, respectively, 
the wave function is  normalized as $\int \vert \psi({\bf r},t) \vert^2 d{\bf r}=1.$  The coefficient 
of the beyond-mean-field  LHY interaction $\gamma_{\mathrm{LHY}}$ is given by \cite{qf1,qf2,blakie}
\begin{align}\label{qf}
\gamma_{\mathrm{LHY}}= \frac{128}{3}\sqrt{\pi a^5} Q_5(\varepsilon_{\mathrm{dd}}),
\end{align}
where  the auxiliary function $ Q_5(\varepsilon_{\mathrm{dd}})$ is given by 
\begin{equation}
 Q_5(\varepsilon_{\mathrm{dd}})=\ \int_0^1 dx(1-\varepsilon_{\mathrm{dd}}+3x^2\varepsilon_{\mathrm{dd}})^{5/2}. 
\end{equation}
This function  can be evaluated as \cite{blakie}
\begin{align}\label{exa}
Q_5(\varepsilon_{\mathrm{dd}}) &=\
\frac{(3\varepsilon_{\mathrm{dd}})^{5/2}}{48}  \Re \left[(8+26\eta+33\eta^2)\sqrt{1+\eta}\right.\nonumber\\
& + \left.
\ 15\eta^3 \mathrm{ln} \left( \frac{1+\sqrt{1+\eta}}{\sqrt{\eta}}\right)  \right], \quad  \eta = \frac{1-\varepsilon_{\mathrm{dd}}}{3\varepsilon_{\mathrm{dd}}},
\end{align}
where $\Re$ denotes the real part.

Equation (\ref{eq.GP3d}) can be reduced to 
the following  dimensionless form by scaling lengths in units of $l = \sqrt{\hbar/m\omega_z}$, time in units of $\omega_z^{-1}$,  angular frequency in units of $\omega_z$,  energy in units of $\hbar\omega_z$
and density $|\psi|^2$ in units of $l^{-3}$
\begin{align}\label{GP3d2}
\mbox i \frac{\partial \psi({\bf r},t)}{\partial t} & =
{\Big [}  -\frac{1}{2}\nabla^2
+U({\bf r}) + 4\pi{a} N \vert \psi({\bf r},t) \vert^2
\nonumber\\ &
+3a_{\mathrm{dd}}  N
\int 
U_{\mathrm{dd}}({\bf R})
\vert\psi({\mathbf r'},t)\vert^2 d{\mathbf r}'   \nonumber \\ 
&+\gamma_{\mathrm{LHY}}N^{3/2}
|\psi({\mathbf r},t)|^3
\Big] \psi({\bf r},t),\\
U({\bf r})&={\textstyle \frac{1}{2}}\left({\omega_x^2}x^2+ {\omega_y^2}y^2+z^2\right),
\label{pot}
\end{align}
where, and in the following, without any risk of confusion, unless otherwise indicated, all variables are scaled 
and represented by the same symbols as the unscaled variables.
 
 Equation (\ref{GP3d2})  
can also be obtained from the variational rule
\begin{align}
i \frac{\partial \psi}{\partial t} &= \frac{\delta E}{\delta \psi^*} 
\end{align}
with the following energy functional (energy per atom)
\begin{align}
E &= \frac{1}{2}\int d{\bf r} \Big[ {|\nabla\psi({\bf r})|^2} +\Big({\omega_x^2}x^2+ {\omega_y^2}y^2+z^2\Big)|\psi({\bf r})|^2\nonumber  \\
&+ {3}a_{\mathrm{dd}}N|\psi({\bf r})|^2 
\left. \int U_{\mathrm{dd}}({\bf R})
|\psi({\bf r'})|^2 d {\bf r'} \right. \nonumber \\
& + 4\pi Na |\psi({\bf r})|^4 +\frac{4\gamma_{\mathrm{LHY}}}{5} N^{3/2}
|\psi({\bf r})|^5\Big]
\end{align}
for a stationary state.

First, a stationary quasi-1D or a quasi-2D 
 dipolar   supersolid state 
is generated  
by solving Eq. (\ref{GP3d2}) by imaginary-time 
propagation. To study the  {dipole-mode} oscillation, we perform real-time propagation with the following space-translated potential 
\begin{align}\label{lintrap}
 U({\bf r})&={\textstyle \frac{1}{2}}\left[{\omega_x^2}(x-x_0)^2+ {\omega_y^2}y^2+z^2\right],
\end{align}
and using the stationary state as the initial function, where $x_0$ is the space translation along the $x$ direction.  In this case a  nondipolar  BEC executes the  simple-harmonic oscillation $x(t)=x_0\cos(\omega_x t)$
along the $x$ direction without any distortion with angular frequency $\omega_x$ indicating superfluidity.

To study the angular scissors-mode oscillation of a quasi-1D supersolid state in the $x$-$y$ plane,  the real-time propagation is executed with the following space-rotated trap
\begin{align}\label{scitrap}
 U({\bf r})&={\textstyle \frac{1}{2}}\left[{\omega_x^2}(x \cos \theta_0 {+} y \sin \theta_0)^2\right.
\nonumber \\ &+ \left. {\omega_y^2}({-} x\sin \theta_0 +y \cos \theta_0 )^2+z^2\right],
\end{align}
employing the stationary state as the initial function, where $\theta_0$ is the angle of rotation of the potential around the polarization $z$ direction. For a sufficiently large asymmetry of the trap in the $x$-$y$ plane a (superfluid) BEC,  { in the Thomas-Fermi regime, obeying the hydrodynamic equations of superfluids,} will execute sustained periodic scissors-mode  oscillation $\theta(t)=\theta_0 \cos(\omega_{\mathrm{th}} t)$
with the frequency $\omega_{\mathrm{th}} = \sqrt{(\omega_x^2+\omega_y^2)}$ \cite{scith,sciex2}. A sustained  angular oscillation with the frequency $\omega_{\mathrm{th}}$ signals superfluidity.
 In the opposite collisionless  regime, distinct  frequencies $|\omega_x\pm \omega_y|$ survive \cite{scith}. Because of the asymmetry of the dipolar interaction, say in the $x$-$z$ plane, it is also possible to have a spontaneous scissors-mode oscillation \cite{sci-sym} of a dipolar BEC  in this plane  with a circularly symmetric trapping potential. The emergent circularly asymmetric dipolar BEC will naturally point along the $z$ direction in this case and, if angularly displaced, can execute  a scissors-mode oscillation in the $x$-$z$ plane. That scissors-mode oscillation is typically different from the present scissors-mode oscillation generated in a circularly asymmetric trap in the $x$-$y$ plane. 

To generate a quasi-1D dipolar supersolid {along the $x$ direction} we need to take
 {$ \omega_x \ll \omega_y , 1$};  to generate a quasi-2D dipolar supersolid {in the $x$-$y$ plane} we will take $\omega_x, \omega_y \ll 1$.  (The angular frequencies are expressed in units of the angular frequency in the $z$ direction, which in dimensionless unit is 1.)
In the case of  {dipole-mode} oscillation we will consider both a
quasi-1D and quasi-2D dipolar supersolid and in the case of
scissors-mode oscillation we will mostly study only a quasi-1D dipolar supersolid. We could not find any sustained scissors-mode oscillation in the case of a quasi-2D dipolar supersolid for any sets of parameters.

\section{Numerical Results}

\label{III}
 
To study the oscillation dynamics  of a  dipolar supersolid
we  solve   partial differential beyond-mean-field  GP  
 equation (\ref{GP3d2}),  numerically, using FORTRAN/C programs \cite{dip} or their open-multiprocessing versions \cite{omp,ompF}, 
employing  the split-time-step Crank-Nicolson
method using the imaginary-time propagation rule  \cite{crank}.
Because of the divergent $1/|{\bf R}|^3$ term in the dipolar potential (\ref{eq.con_dipInter}),
 it is problematic to treat numerically the nonlocal dipolar interaction integral in the beyond-mean-field model (\ref{GP3d2}) in configuration space. To circumvent the problem, this term is evaluated in the momentum $\bf k$
 space by a Fourier transformation using a convolution identity as \cite{dip}
 \begin{align}
 \int d{\bf r'}U_{\mathrm{dd}}({\bf R})n({\bf r'})=\int \frac{d \bf k}{(2\pi)^3}e^{-i\bf k\cdot r} \widetilde V_{\mathrm{dd}}({\bf k})  \widetilde n({\bf k}),
 \end{align}
 where $n({\bf r})\equiv |\psi({\bf r})|^2$;   $\widetilde V_{\mathrm{dd}}({\bf k})$  and $\widetilde n({\bf k})$
 are respective Fourier transforms. 
 This is advantageous numerically due to the smooth behavior of this term in momentum space. 
 The Fourier transformation of the dipolar potential  $\widetilde V_{\mathrm{dd}}({\bf k})$ can be found analytically \cite{dip} enhancing the accuracy of the numerical procedure.
 After solving the problem in momentum space, a backward  Fourier transformation provides the desired solution in configuration space.

  For the appearance of { a supersolid droplet-lattice} we need a strongly dipolar atom with 
  {$a_{\mathrm{dd}}>a$}   \cite{2d3}. The system becomes repulsive for 
 { $a_{\mathrm{dd}}<a$,}
{while the system is necessarily a superfluid,}   and no droplets can be formed. Instead of presenting  results only  in dimensionless
units, we also
relate our results to the recent experimental  \cite{2d2,sci-y} and related theoretical \cite{2d4} studies   on
supersolid formation in a dipolar BEC of $^{164}$Dy atoms. Although,  $a_{\mathrm{dd}}=130.8a_0$ for  $^{164}$Dy atoms, where $a_0$  is the Bohr radius,
we have a certain flexibility in fixing the scattering length $a$, as the scattering length can be modified by 
the Feshbach resonance technique by manipulating an external electromagnetic field.
  As in our previous studies \cite{luis,luis2}, 
 we take $a=85a_0$, which is close to its experimental estimate $a=(92\pm 8)a_0$  \cite{scatmes}.
 With the reduction of contact repulsion, 
 this choice has the advantage of slightly increasing the net attraction, which will facilitate the formation of the dipolar droplets.
 Consequently, { we use   {$a=85a_0$} in all  calculations of this paper}.  Other studies on quantum droplets in a quasi-2D dipolar BEC used nearby values of scattering lengths, e.g.,  
$a=88a_0$,   \cite{2d2,2d4} and $a=70a_0$   \cite{blakieprl} $-$ always smaller than its experimental estimate to facilitate the formation of droplets and droplet lattice.

\begin{figure}[t!]
\begin{center}
\includegraphics[width=.49\linewidth]{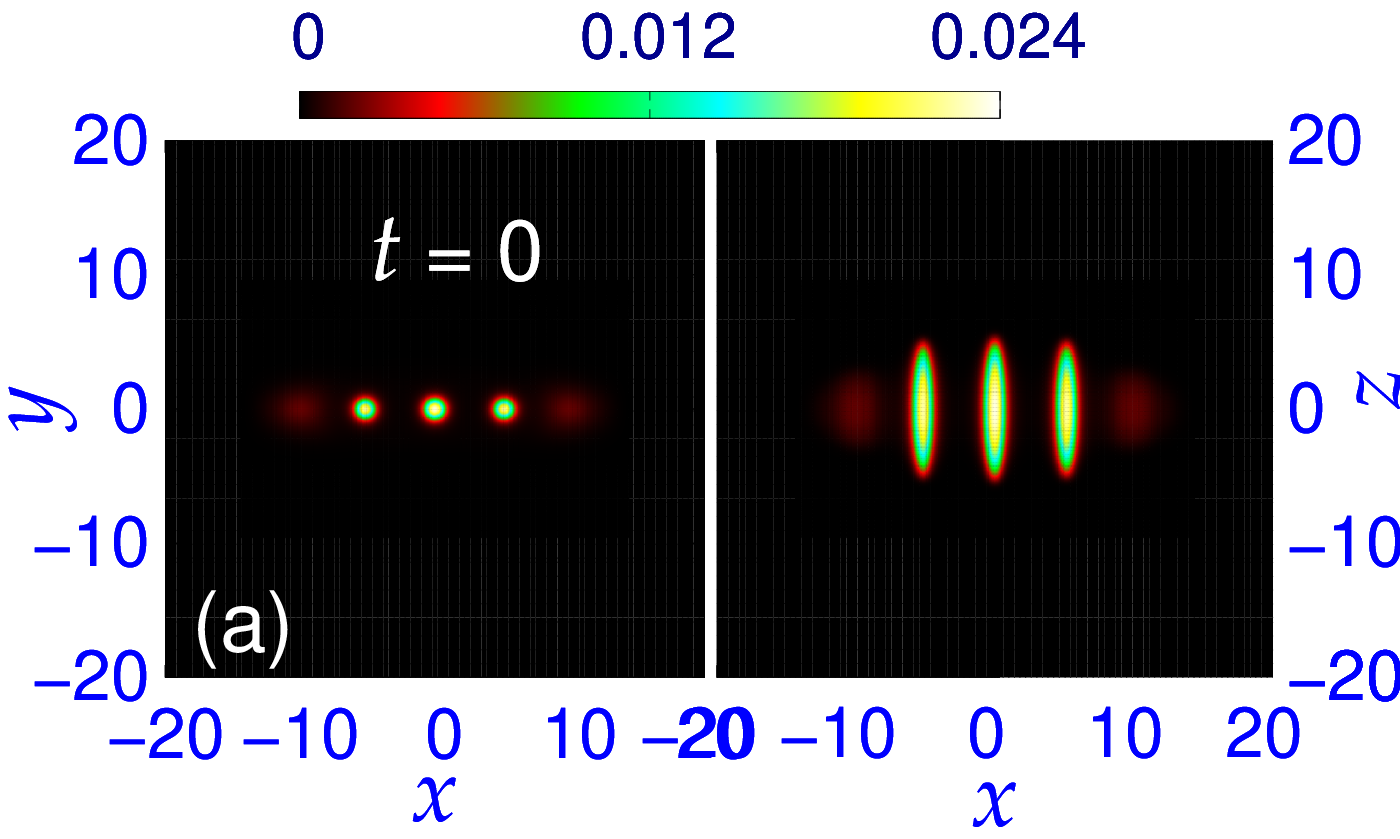}
\includegraphics[width=.49\linewidth]{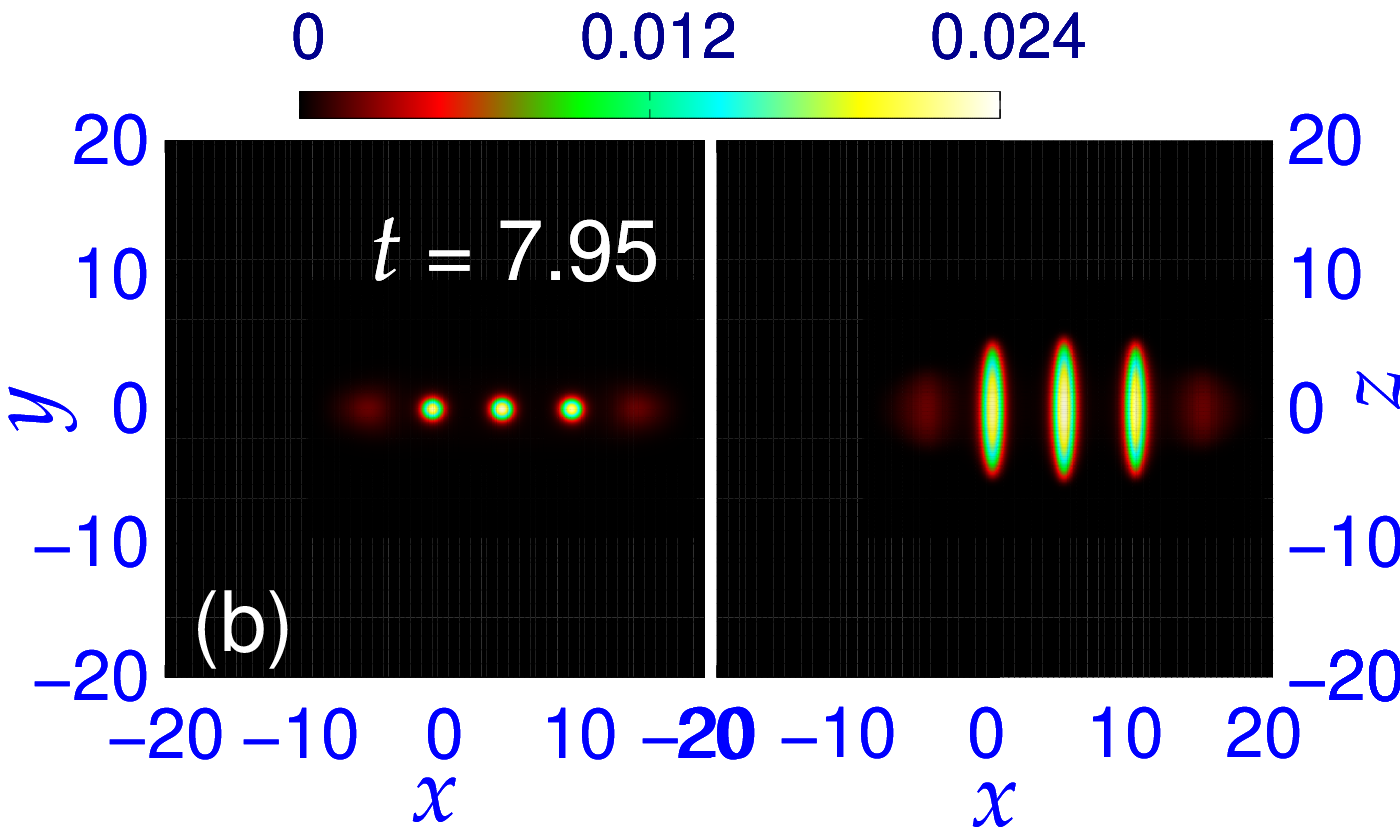}
\includegraphics[width=.49\linewidth]{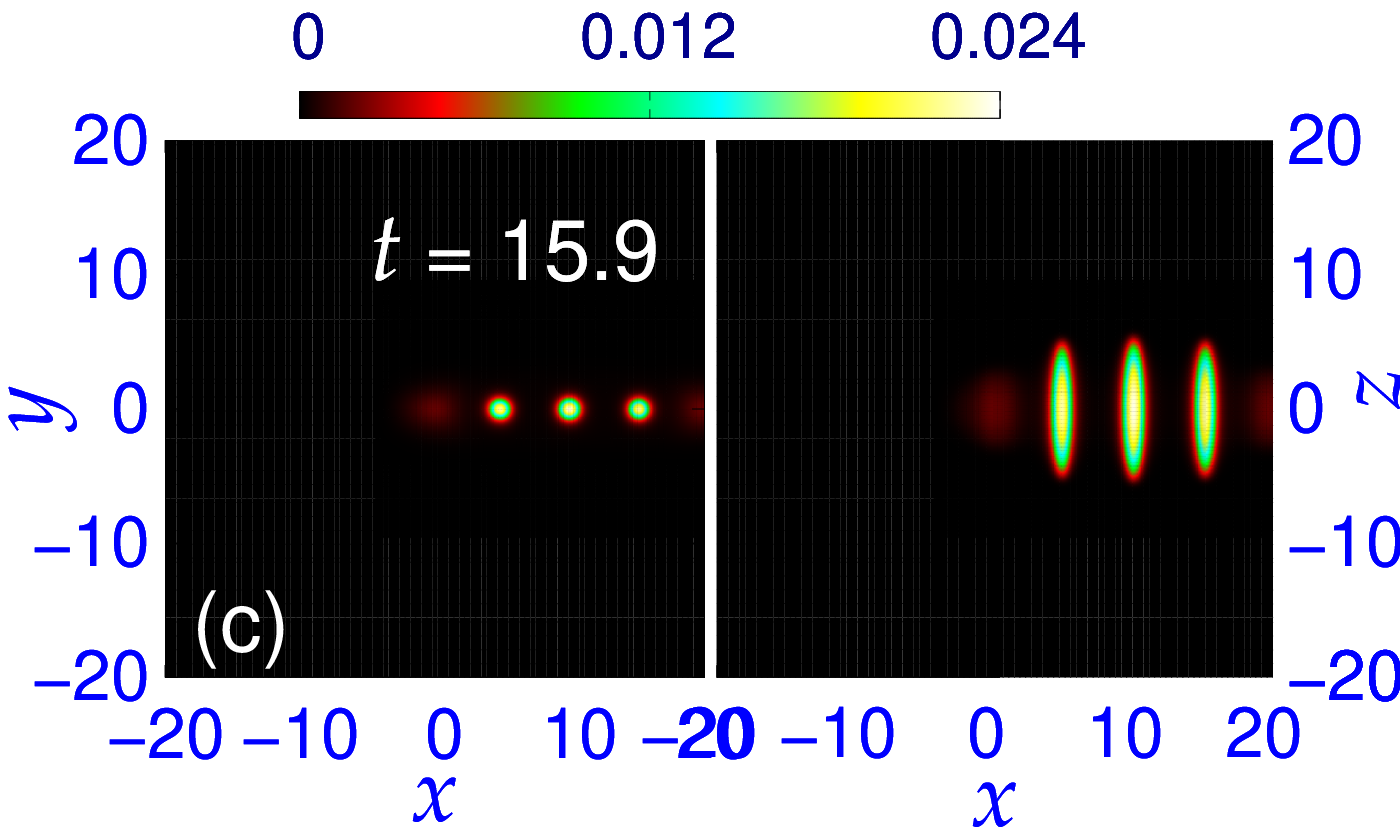}
\includegraphics[width=.49\linewidth]{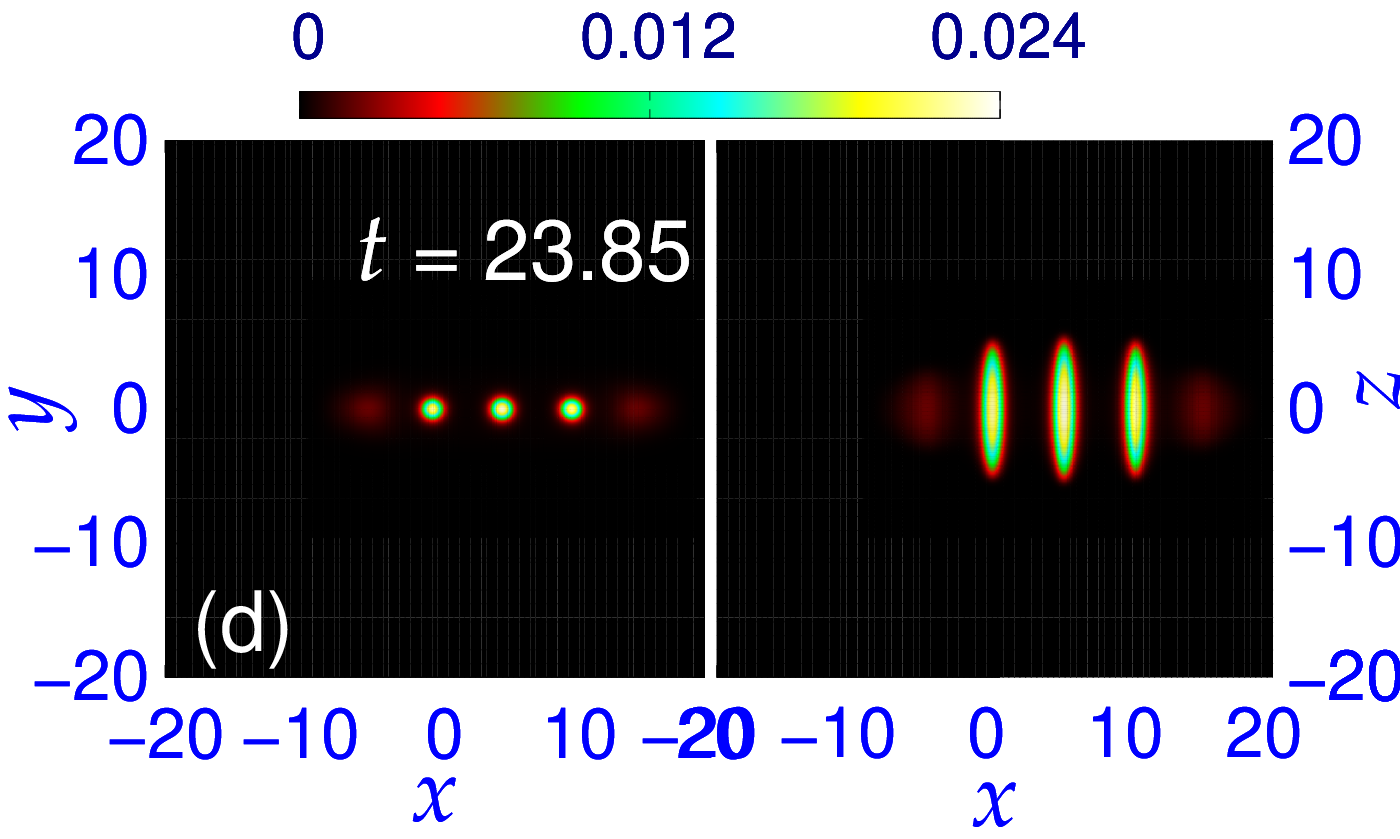}
\includegraphics[width=.49\linewidth]{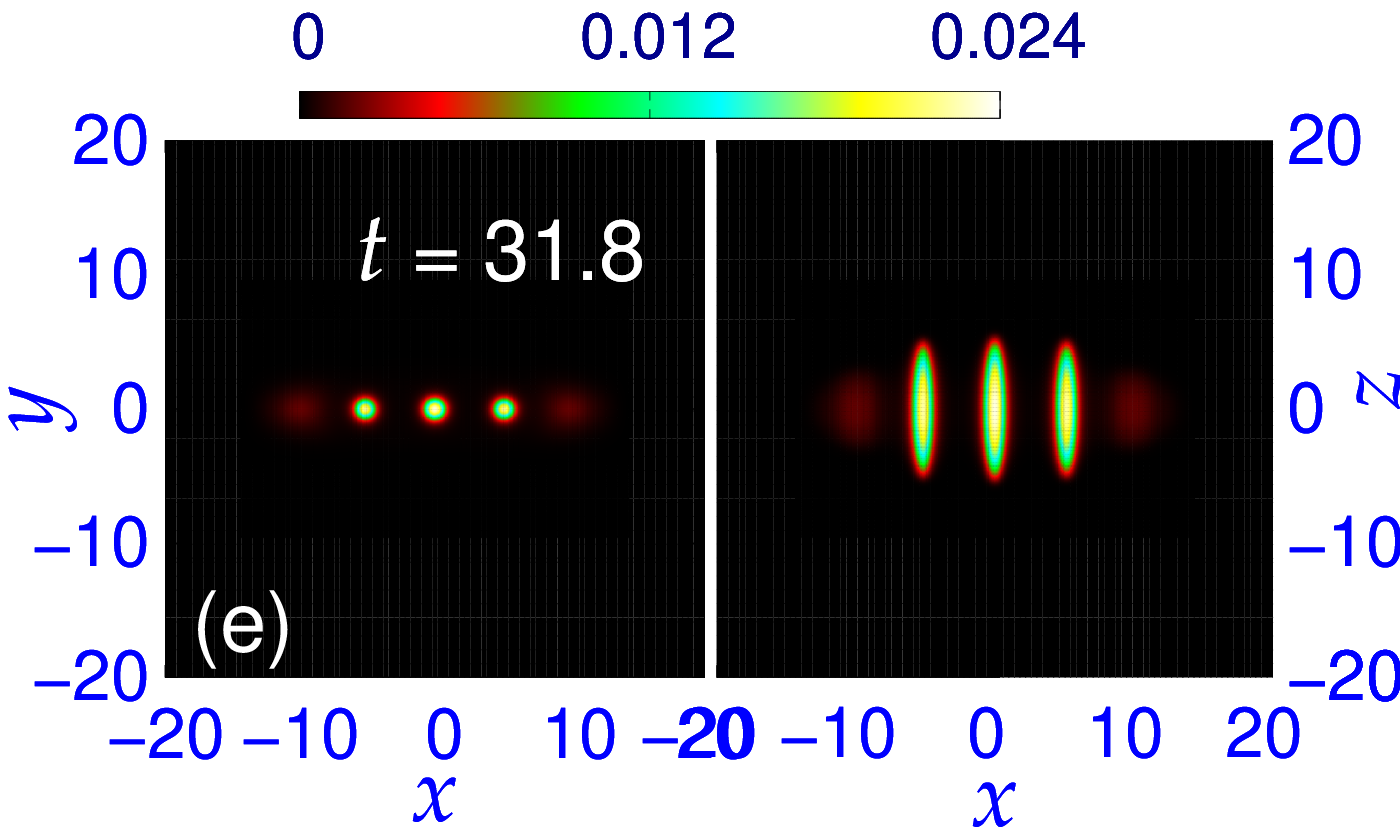}
\includegraphics[width=.49\linewidth]{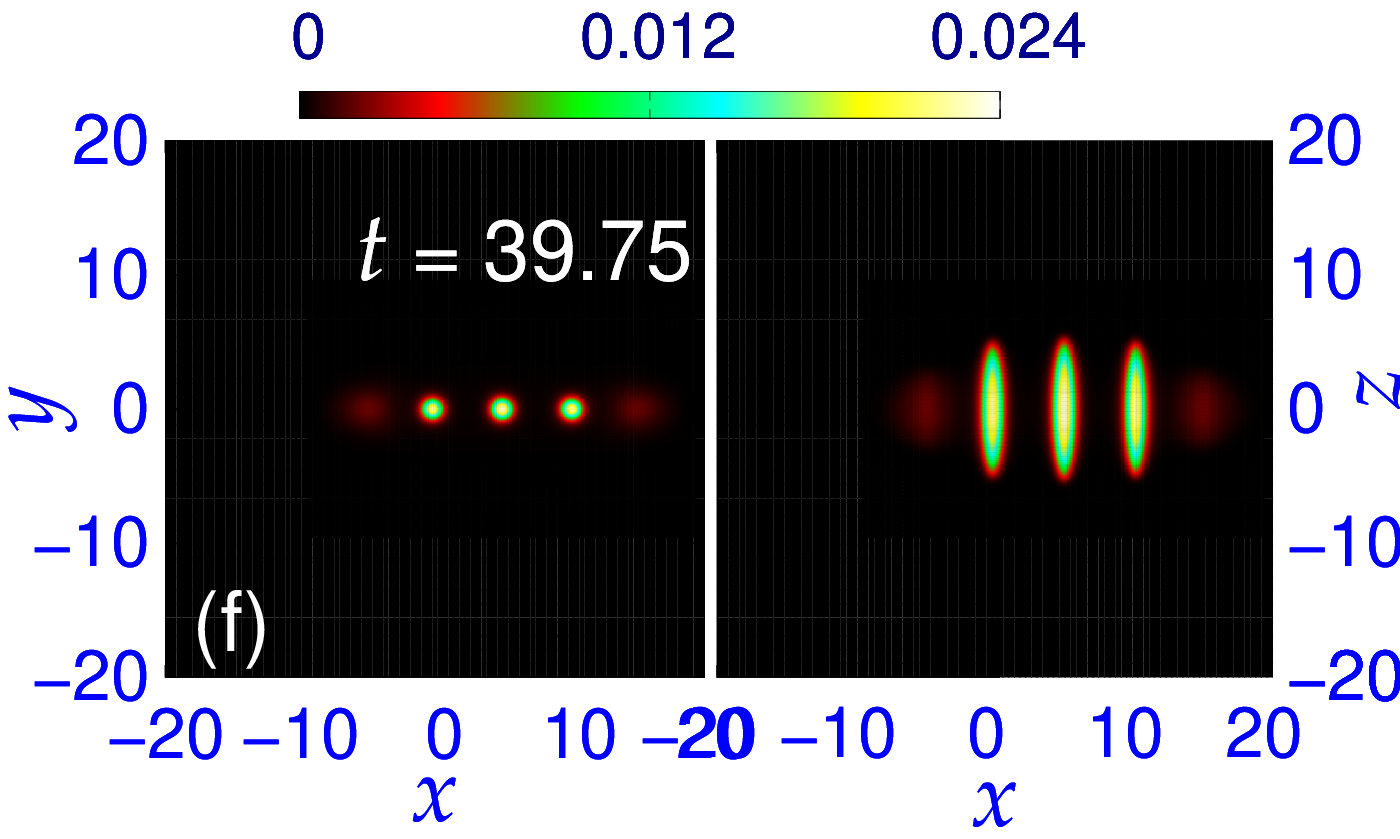}

\caption{ Contour plot of density $|\psi(x,y,0)|^2$ (left side of each {panel})  and $|\psi(x,0,z)|^2$ (right side of each {panel})
of the quasi-1D three-droplet supersolid  of $N=20000$ $^{164}$Dy atoms 
executing linear  {dipole-mode oscillation in trap (A)}
at times (a)
$t=0$, (b) $t= 7.95$, (c) $t=15.9$, (d) $t=23.85$, (e) $t=31.8$, (f) $t=39.75$.  {The trap frequencies are $\omega_x =33/167, \omega_y = 110/167, \omega_z=1.$}
 Displayed results in all figures [except Fig. \ref{fig5}(a)] are dimensionless. {The unit of length is $l=0.6075$ $\mu$m and unit of time  $0.953$ ms.}
}
\label{fig1} 
\end{center}
\end{figure}

In this study  the trap frequencies along $x$ and $z$ directions  are 
taken as (A) $\omega_x=33/167$,   $\omega_z=1$,   as in a recent experimental \cite{2d2} and related theoretical \cite{2d4} investigations on   hexagonal-lattice crystallization of droplets. 
For dysprosium atoms   $m(^{164}$Dy)    $\approx 164 \times 1.66054\times 10^{-27}$ kg, $\hbar =  1.0545718 \times  10^{-34}$ m$^2$ kg/s, $\omega_z = 2\pi \times 167 $  Hz, consequently, the unit of length $l=\sqrt{\hbar/m\omega_z}= 0.6075$  $\mu$m.  To find the dependence of scissors-mode-oscillation 
frequency on the trapping frequency, we also considered the frequencies (B)  $\omega_x=23/90$ and $\omega_z=1$, as in a recent experimental study of scissors-mode oscillation of a quasi-1D dipolar supersolid \cite{sci-y}; in that case  $l=0.8275$ $\mu$m. { In both cases  the trap frequency along the $y$ direction $\omega_y$ will be varied to generate an appropriate  quasi-1D or a quasi-2D trap.  
 }

 \subsection{ {Dipole-mode} oscillation of a quasi-1D and quasi-2D supersolid}

 To prepare a quasi-1D dipolar supersolid for the investigation of dynamics,
we consider  20000 $^{164}$Dy atoms in 
 the quasi-1D trap (A)  with $\omega_x =33/167$, $\omega_y=110/167$ and $\omega_z=1$.    The dipolar BEC crystallizes in a three-droplet state along the $x$ axis.  The  converged final state in this case can be obtained by imaginary-time simulation using an initial  Gaussian wave function.  However, the convergence is quicker if we use an 
analytic wave function for  a few droplets (3 or 5) periodically arranged  along the $x$ direction with a fixed mutual 
separation and symmetrically placed around the occupied $x=0$ site as in Ref. \cite{luis} and we will take such an initial state in the present study.   A contour plot of the $z=0$ and $y=0$  sections of the 3D density 
$|\psi(x,y,0)|^2$ (left side) and  $|\psi(x,0,z)|^2$ (right side) is  shown in Fig. \ref{fig1}(a) with 3 droplets placed symmetrically around $x=0$ (a parity-symmetric state).

\begin{figure}[t!]
\begin{center}
\includegraphics[width=.49\linewidth]{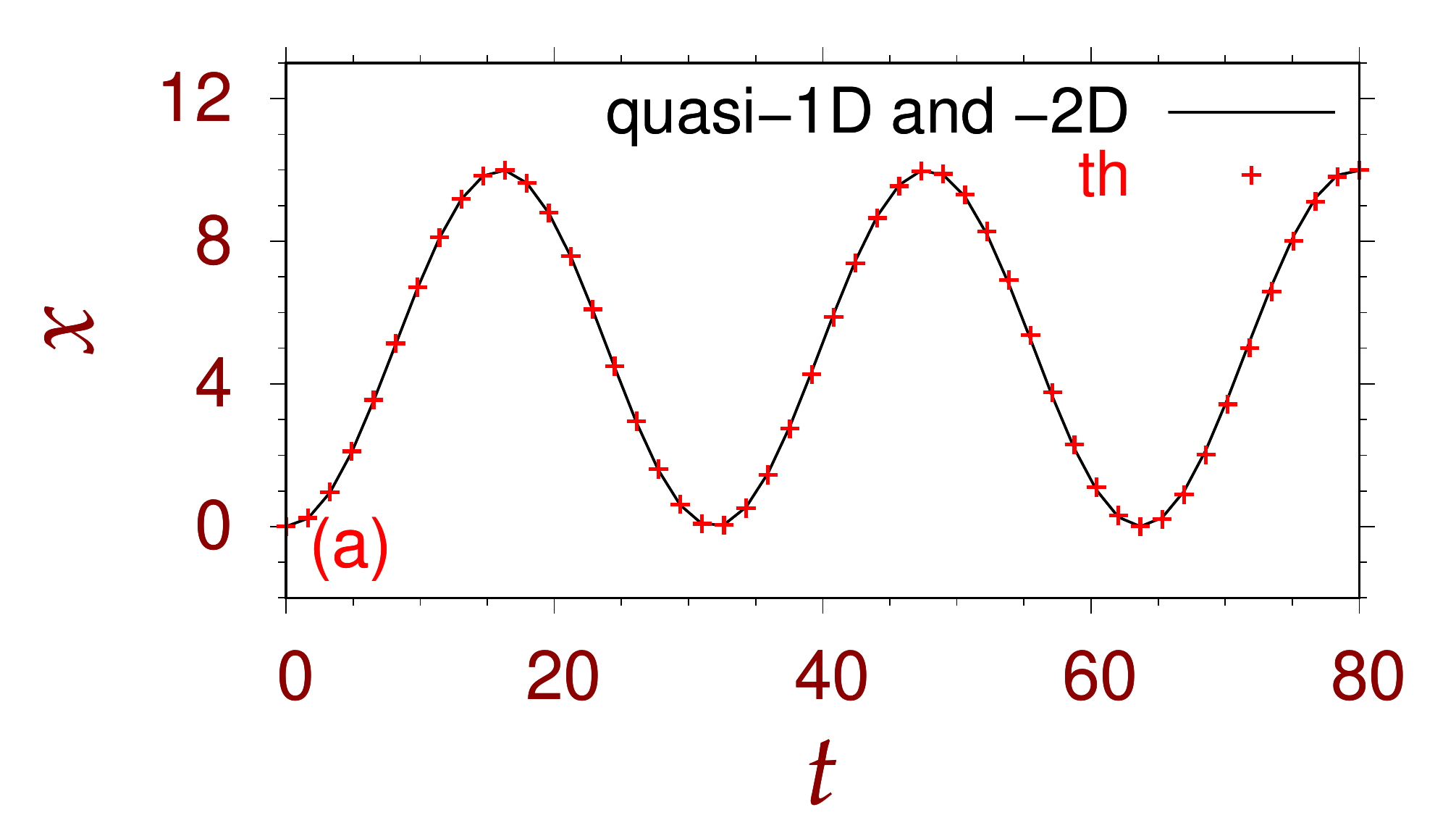}
\includegraphics[width=.49\linewidth]{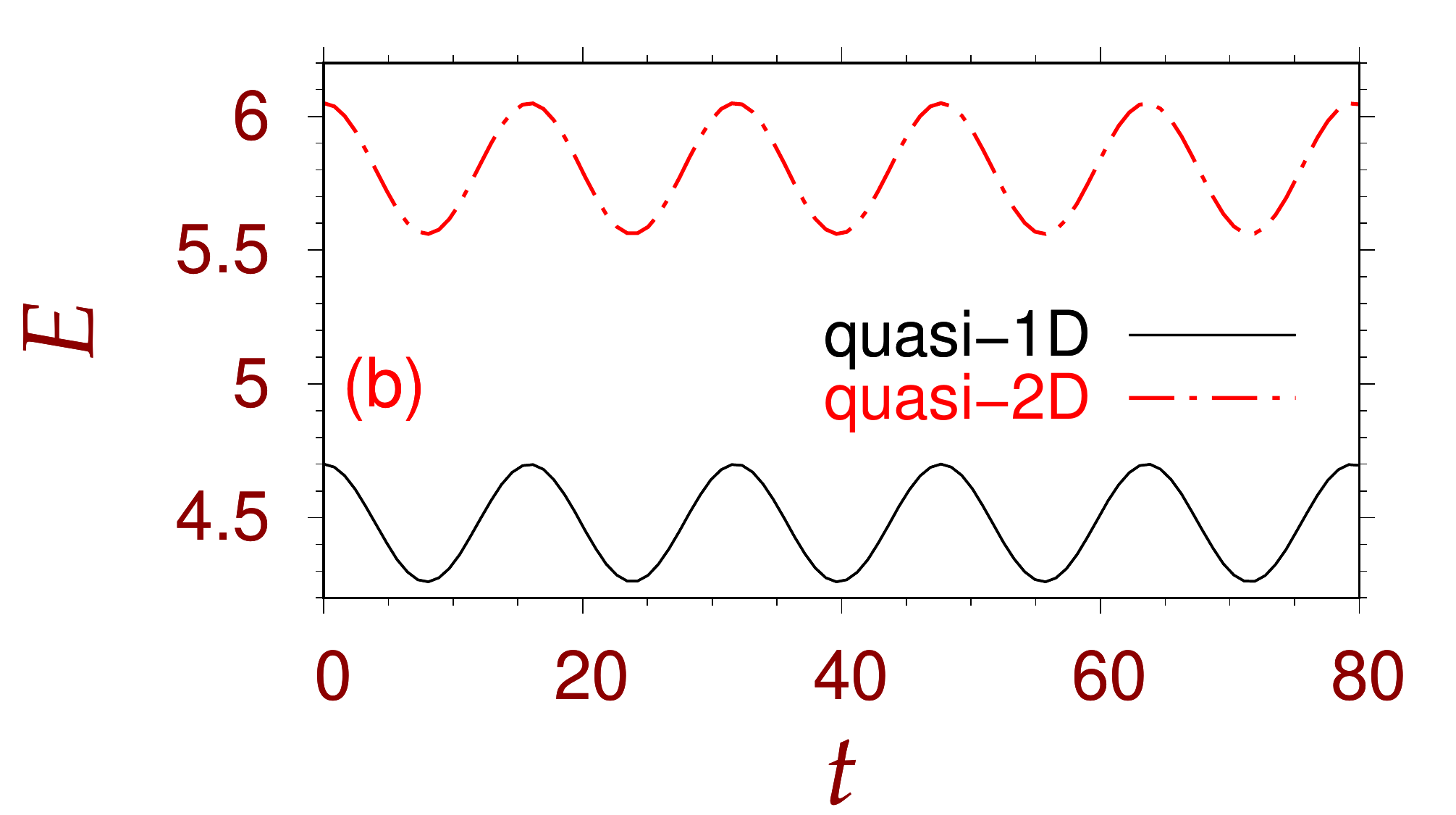}
\caption{(a) Linear displacement $x$ { of the central droplet} versus time $t$ of a quasi-1D three-droplet supersolid of $N=20000$ $^{164}$Dy atoms 
or of a quasi-2D nine-droplet supersolid of $N=60000$ $^{164}$Dy atoms 
 executing  {dipole-mode}  oscillation along $x$ direction
as obtained by real-time propagation   fitted to the theoretical prediction  $\cos(\omega_xt)$ (th).  
 The oscillation is started by  a linear displacement  of $x_0=5$ of  trap (A) at $t=0$, viz. Eq. (\ref{lintrap}). (b) Energy $E$ versus time $t$  during this  {dipole-mode} oscillation.
{ For quasi-1D case, the angular frequency $\omega_y=110/167 $ and for quasi-2D case  $\omega_y =33/167$. Other parameters are  $\omega_x = 33/167$,      $\omega_z=1$,  $a=
85a_0/l$ and $l=0.6075$.} 
}
\label{fig2} 
\end{center}
\end{figure}

  To study the   {dipole-mode oscillation}, we consider the above-mentioned  quasi-1D 
 three-droplet dipolar supersolid  in trap (A) \cite{2d2,2d4}, 
 displaced along the $x$ direction through a distance of $x_0=5$, viz. Eq. (\ref{lintrap}).
 { For both quasi-1D and quasi-2D  dipole-mode oscillations (studied in the following),
 the initial configuration is the stationary state obtained by imaginary-time propagation, and the  dynamics is studied by real-time simulation replacing the original symmetric trap 
 (\ref{pot}) by the 
  displaced trap (\ref{lintrap}) with $x_0=5$ at $t=0$.}
  Due to the linear displacement of the trap along the $x$ direction, the dipolar supersolid will execute sustained  {dipole-mode} oscillation along the $x$ direction with an amplitude of 5.
  In Fig. \ref{fig2}(a) we compare the time evolution of position $x$ 
{of the central droplet}
 with its theoretical prediction of periodic oscillation with the trap frequency $\omega_x$. The present period of oscillation $T=31.8$ compares well with the theoretical period $T\equiv 2\pi /\omega_x =31.7967.$
  The energy of the oscillating supersolid also executes a steady simple-harmonic oscillation as shown in Fig. \ref{fig2}(b).  The frequency of energy oscillation is double that of the frequency of position oscillation.
   { The dependence  of the energy of the oscillating supersolid 
   with time is important as
   it  can be calculated much more accurately than the position of the supersolid
   and any deviation from the expected simple harmonic oscillation of energy signals a breakdown of the expected  dipole-mode and scissors-mode oscillations $-$ indicating either a distortion of the supersolid or a destruction/absence of superfluidity during oscillation
   or both, viz. Fig. \ref{fig9}.}

 The linear   {dipole-mode} oscillation of a quasi-1D dipolar supersolid along the $x$ direction
 is more explicitly illustrated in terms of a contour plot of densities $|\psi(x,y,0)|^2$ (left side) and $|\psi(x,0,z)|^2$ (right side) as displayed in Fig. \ref{fig1}
at times (a) $t=0$, (b) $t= 7.95$, (c) $t=15.9$, (d) $t= 23.85$, (e) $t=31.8$, and (f) $t= 39.75$. 
At (a) $t=0$ the center of the supersolid lies at $x=0$, at (b) $t= 7.95$  it moves to the center of the displaced trap at $x=5$, and at (c) $t=15.9 $ it moves to the position of largest displacement at $x=10$. After that the supersolid turns around  and passes through the center of  the displaced trap again at  (d)  $t= 23.85$ to the initial equilibrium position at (e) $t=31.8$ at the end of a complete period. Then the system starts to repeat
the same cycle again passing through the center of the displaced trap at (f) $t= 39.75 $. This  {dipole-mode} oscillation is found to be simple harmonic with  the present amplitude of 5.

\begin{figure}[t!]
\begin{center}
 \includegraphics[width=.49\linewidth]{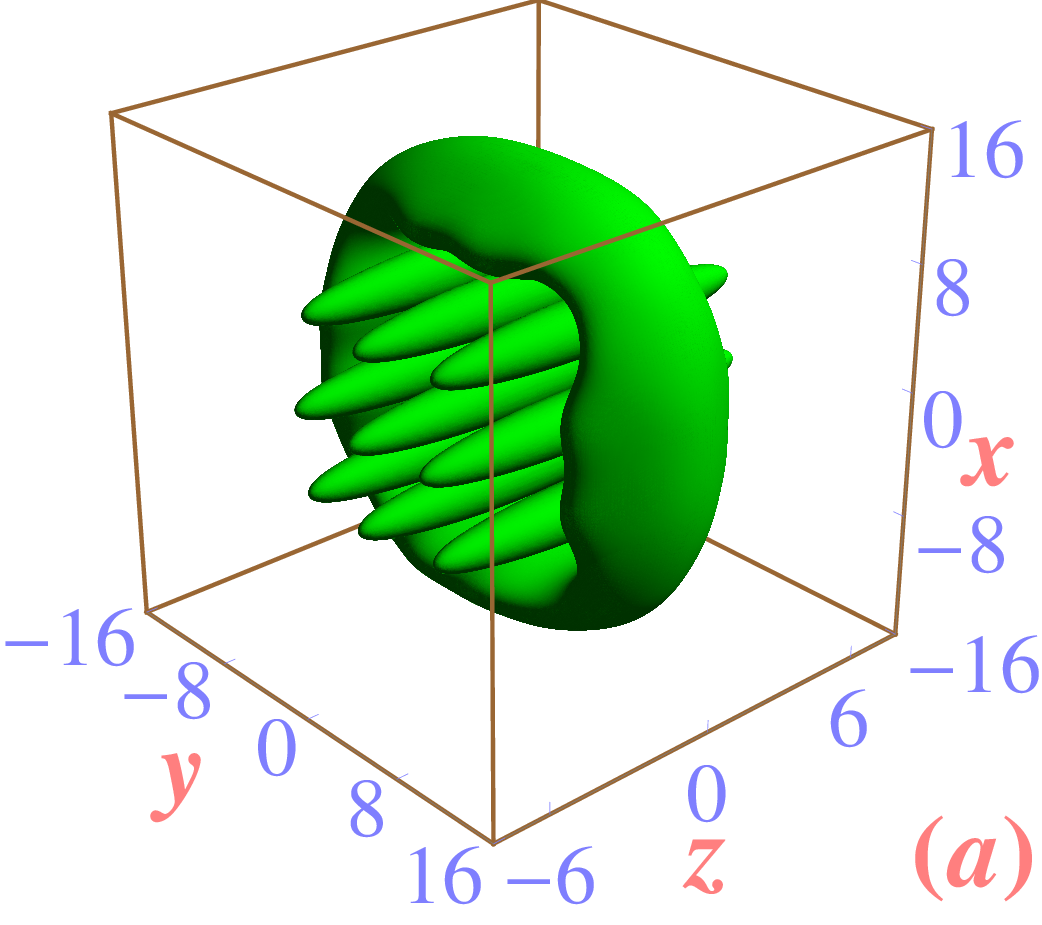}
\includegraphics[width=.49\linewidth]{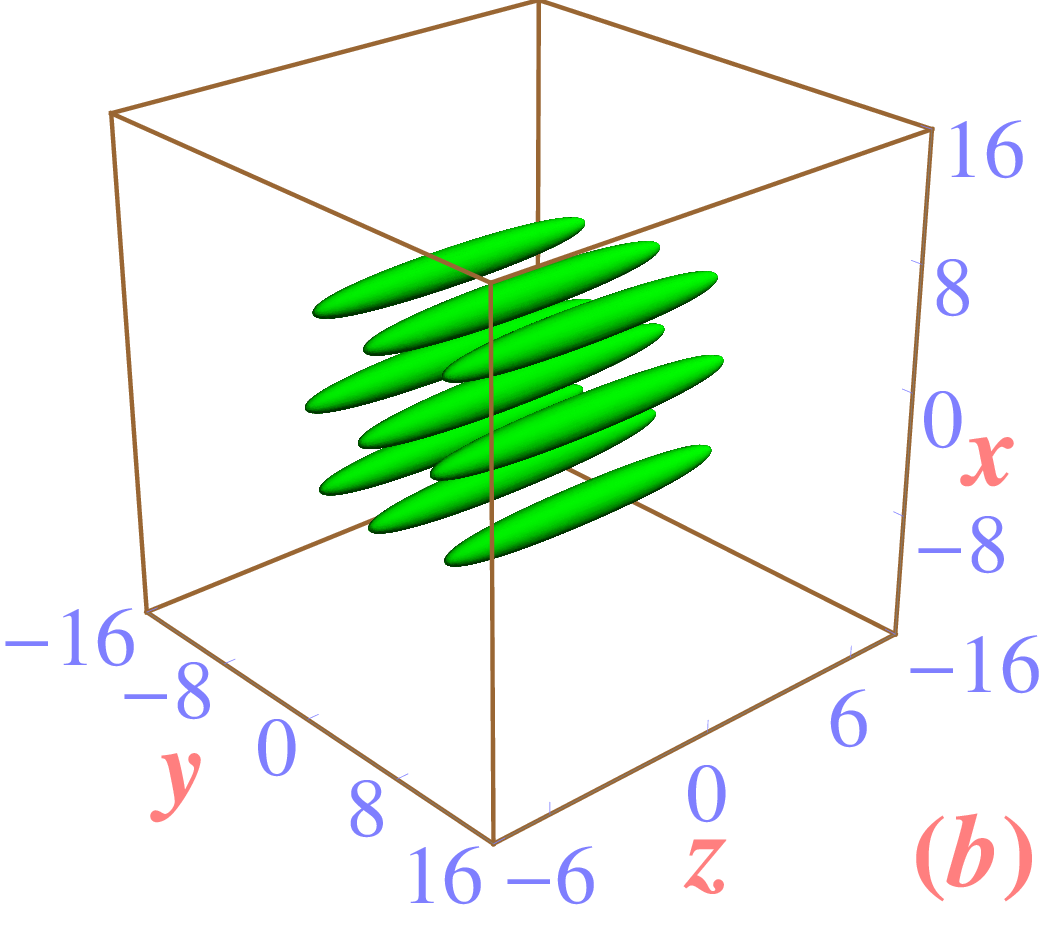}

\caption{Three-dimensional isodensity plot of  $|\psi(x,y,z)|^2$  
of a quasi-2D   nine-droplet (square-lattice) supersolid of $N=60000$  $^{164}$Dy atoms  in { trap (A)} with $\omega_x=\omega_y=33/167,  \omega_z=1$
for the value of density on  contour of (a) 0.0001 and (b) 0.0005.   
}
\label{fig3} 
\end{center}
\end{figure}

\begin{figure}[t!]
\begin{center}
\includegraphics[width=.325\linewidth]{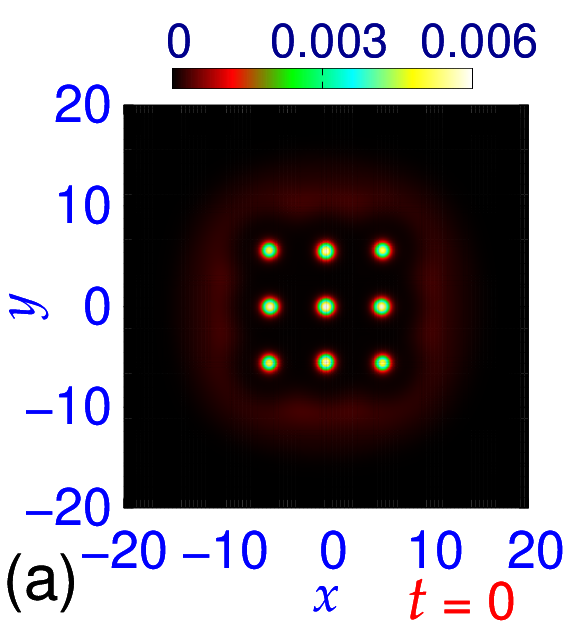}
\includegraphics[width=.325\linewidth]{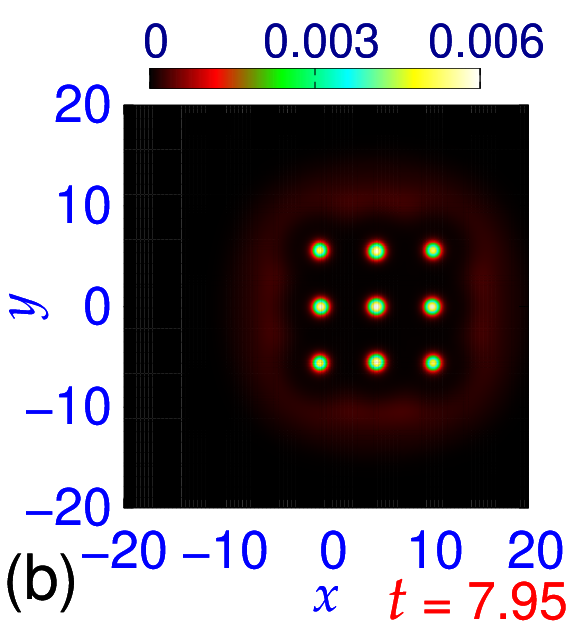}
\includegraphics[width=.325\linewidth]{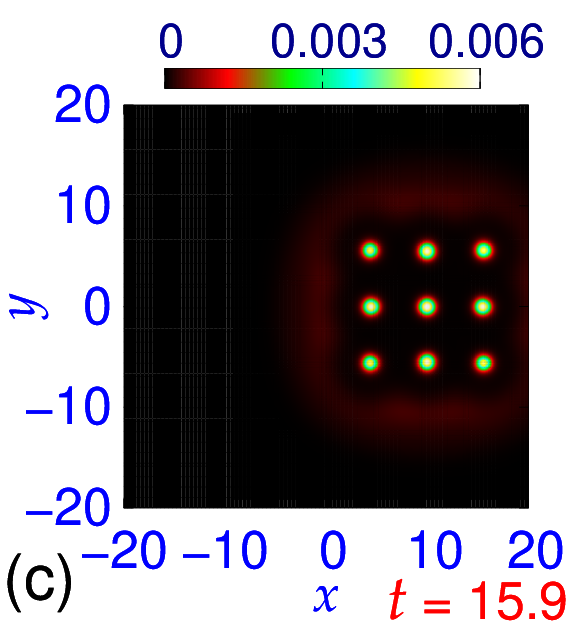}
\includegraphics[width=.325\linewidth]{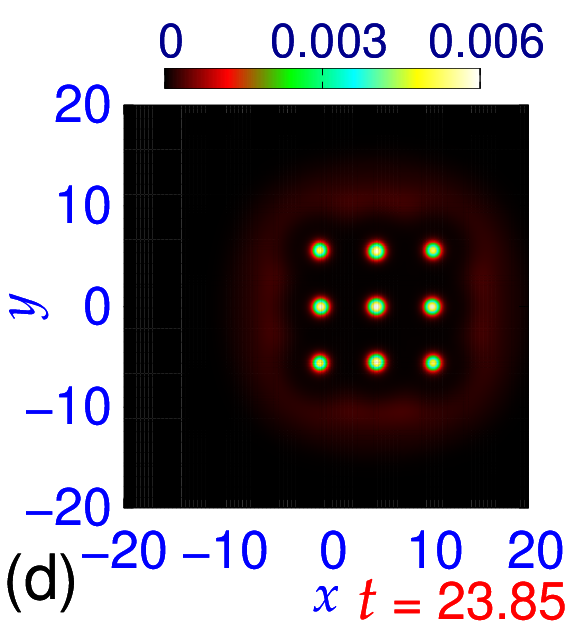}
 \includegraphics[width=.325\linewidth]{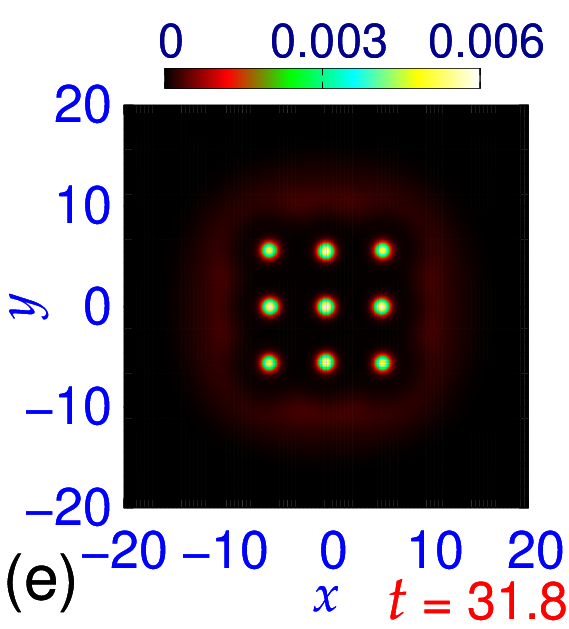}
\includegraphics[width=.325\linewidth]{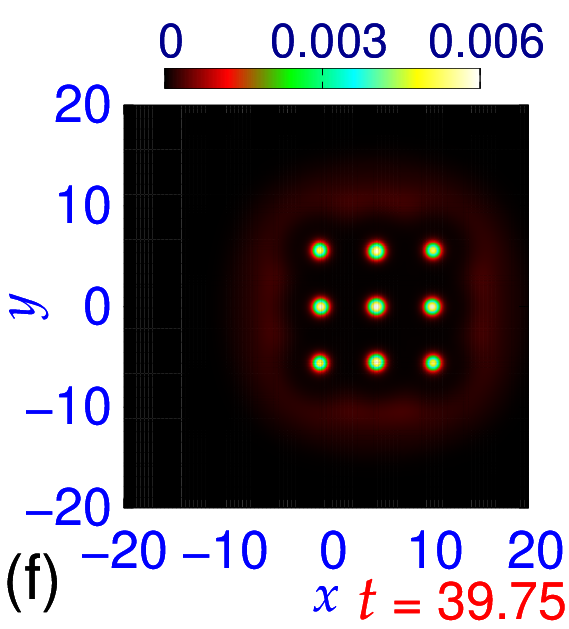}
 
\caption{ Contour plot of density $|\psi(x,y,0)|^2$ 
of the quasi-2D nine-droplet supersolid  of $N=60000$ $^{164}$Dy atoms of Fig. \ref{fig3}
executing  {dipole-mode}  oscillation { in trap (A)}
at times (a)
$t=0$, (b) $t=7.95$, (c) $t=15.9$, (d) $t=23.85$, (e) $t=31.8$, (f) $t=39.75$. {  Trap parameters are 
$\omega_x=\omega_y =33/167, \omega_z=1.$}
}
\label{fig4} 
\end{center}
\end{figure}

Next we consider the linear  {dipole-mode} oscillation of a quasi-2D nine-droplet supersolid of 60000 $^{164}$Dy atoms arranged on a square lattice
in  trap (A) with  angular frequencies $\omega_x=\omega_y =33/167$, and $\omega_z =1$.  The 3D isodensity  plot  of $|\psi(x,y,z)|^2$ of this supersolid is shown in Figs. \ref{fig3}(a)-(b) for densities on contour (a) 0.0001 and (b) 0.0005. A background atom cloud surrounding the square-lattice arrangement of nine droplets can be seen clearly in Fig. \ref{fig3}(a). For a large density on contour in Fig. \ref{fig3}(b) the low-density 
background atom cloud is not visible and a perfect square-lattice arrangement of droplets can be seen. The high-density droplets in a supersolid are embedded in the low-density atom cloud and thus the whole supersolid is phase coherent, which is responsible for frictionless flow and transportability of the supersolid.
Similar background atom cloud also exists in a quasi-1D dipolar supersolid.
The  {dipole-mode} oscillation of the quasi-2D nine-droplet dipolar supersolid  is initiated by displacing the trap through a distance of $x_0=5$ units, viz. Eq. (\ref{lintrap}), and studied by real-time propagation using the stationary wave-function as the initial state.
The time evolution of the position of this quasi-2D supersolid is the same as the quasi-1D supersolid as shown in Fig. \ref{fig2}(a). The time evolution of the energy of this quasi-2D supersolid  is distinct from 
that of the quasi-1D supersolid, viz. Fig. \ref{fig2}(b), although both are controlled by the axial trap frequency in the   $x$ direction $\omega_x$.   The  {dipole-mode} oscillation in the $x$ direction  is better illustrated by snapshots of contour plot of density $|\psi(x,y,0)|^2$ in the $x$-$y$ plane at different times (a) $t=0$, (b) $t=7.95$, (c) $t=15.9$, (d) $t=23.85$, (e) $t=31.8$ and (f) $t=39.75$ as displayed in Fig. \ref{fig4}. The nine-droplet  supersolid starts the oscillation in (a), passes through the position of the minimum of trapping potential at $x=5$ in (b) at $t=7.95$ to  the position of maximum displacement $x=10$ in (c) at $t=15.9$. Then it turns around, passes through the position  $x=5$ in (d) at $t=23.85$ to the initial position $x=0$  in (e) at $t=31.8$, and repeats the same dynamics.   
Although not explicitly demonstrated in this paper, similar oscillation of a hexagonal supersolid was also found.  Sustained  {dipole-mode} oscillation without distortion of both the quasi-1D and quasi-2D dipolar supersolids 
guarantee superfluidity and robustness of the crystalline structure.

 \subsection{Scissors-mode oscillation of a quasi-1D supersolid}

\begin{figure}[t!]
\begin{center}
\includegraphics[width=.49\linewidth]{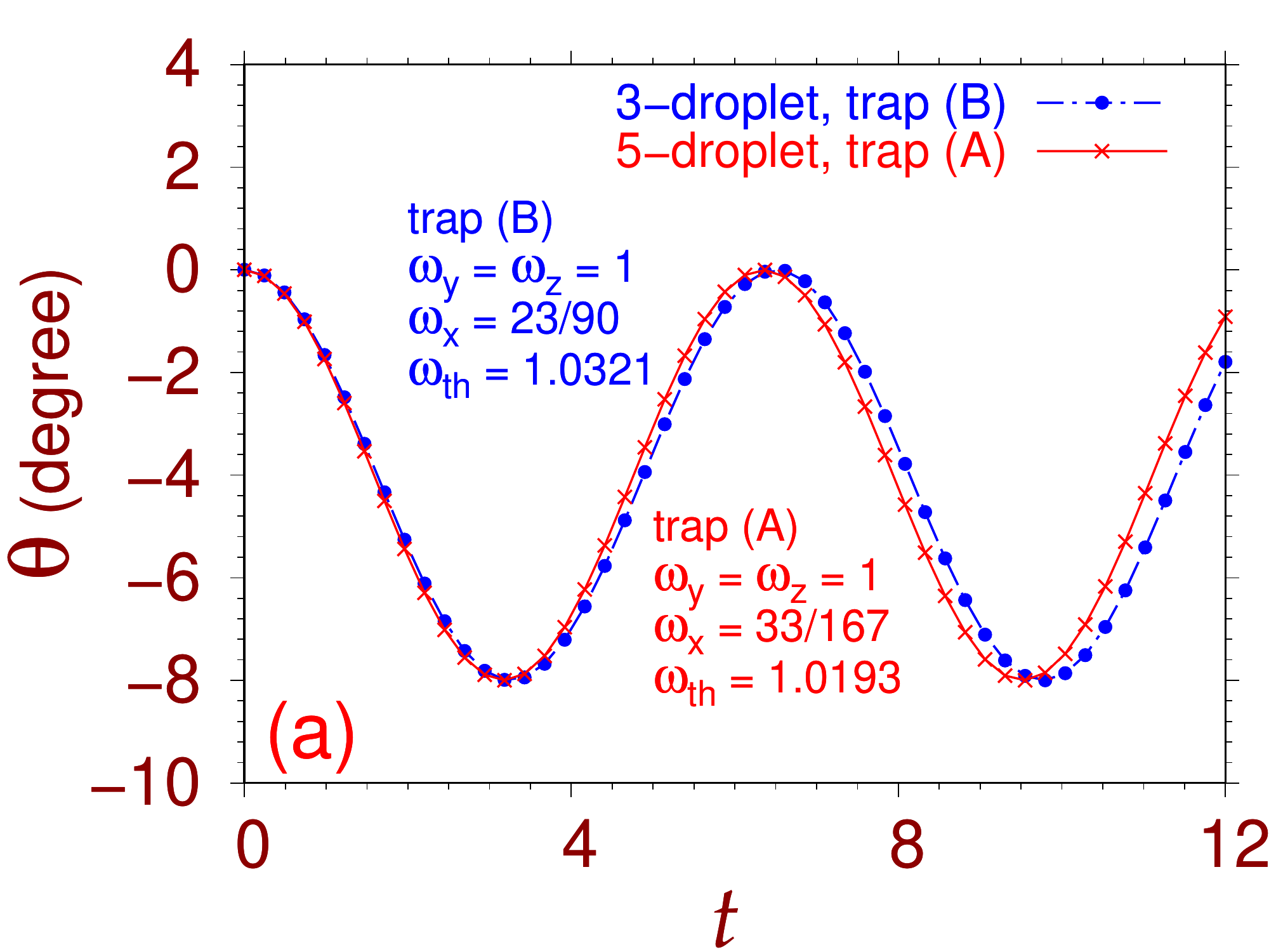}
\includegraphics[width=.49\linewidth]{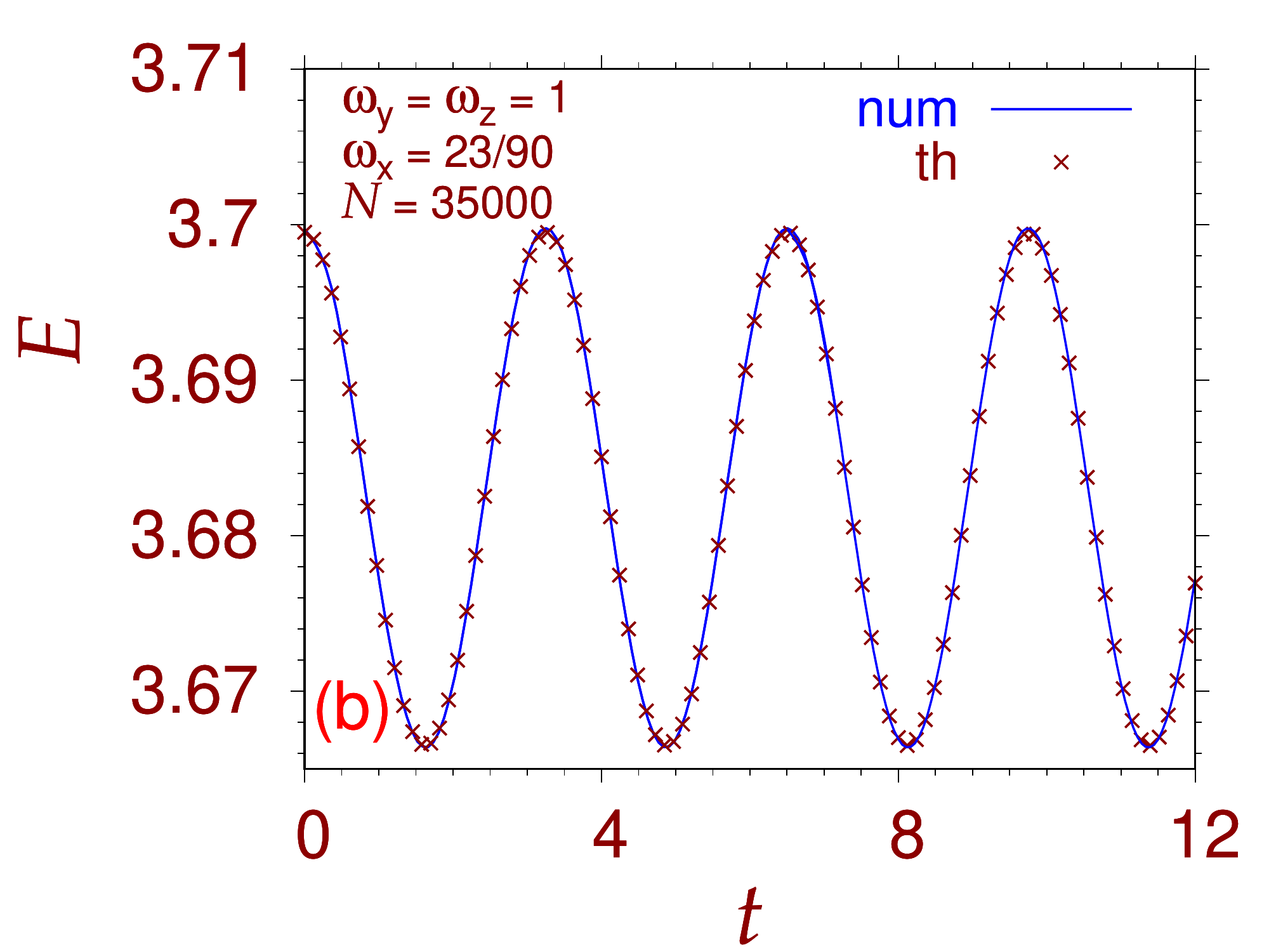}
\caption{(a) Angular displacement $\theta$ versus dimensionless time $t$ of a  quasi-1D three-droplet (five-droplet)
 supersolid of $N=35000$ (40000) $^{164}$Dy atoms  executing scissors-mode oscillation in trap (B) (A) 
fitted to the sinusoidal oscillation
$\cos(0.935\omega_{\mathrm{th}} t)$  $(\cos(0.971\omega_{\mathrm{th}} t))$  \cite{scith,sciex2}.  
 The oscillation is started by giving a rotation of $\theta_0 =-4\degree$ of the trapping potential at $t=0$, viz. Eq. (\ref{scitrap}). (b) Energy $E$ versus time $t$ plot during the scissors-mode oscillation of the three-droplet supersolid
 in trap (B) 
fitted to  the sinusoidal oscillation
$\cos(2\times 0.935\omega_{\mathrm{th}} t)$.  {For trap (A) $\omega_x=33/167, \omega_y=\omega_z=1$, and for trap (B) $\omega_x=23/90, \omega_y=\omega_z=1$. In trap (B) the unit of length is $l=0.8275.$}
 }
\label{fig5}   
\end{center}
\end{figure}

We consider the quasi-1D dipolar supersolid of a few (three or five) droplets  
in 
 trap (A)    $\omega_x =33/167, \omega_z=1$,  and trap (B)  $\omega_x =23/90$, 
    $\omega_z=1$. 
    In both cases we study the variation of the scissors-mode-oscillation frequency with a variation of $\omega_y$. { By varying $\omega_y$ from a small ($\omega_y \ll 1 $)  to a large ($\omega_y \gg 1$) value
 we will pass from a quasi-2D trap to a quasi-1D trap and study scissors-mode oscillation of a quasi-1D supersolid in both types of trap.}
  { The initial configuration in all calculations   is the stationary state obtained by imaginary-time propagation in the appropriate trap, and   
    the angular scissors-mode oscillation is started  by rotating   the spatially-asymmetric  harmonic trap in the $x$-$y$ plane (\ref{pot})
    around the $z$ direction in counter clockwise sense at $t=0$ 
    through an angle $\theta_0=-4\degree$  at $t=0$, viz. (\ref{scitrap}), 
    and  the subsequent dynamics is studied by real-time simulation.}
  Due to the strong spatial asymmetry ($\omega_y \gg \omega_x $) of the trap in the $x$-$y$ plane the dipolar supersolid will execute sustained scissors-mode oscillation \cite{scith,sciex,sci-x} around the $z$ direction. 
  A reasonably large  spatial asymmetry of trap in the $x$-$y$ plane is necessary for a sustained scissors-mode oscillation \cite{scith}.
  {The theoretical frequency of this oscillation for a large superfluid BEC with a TF distribution of matter 
  is $\omega_{\mathrm{th}}=\sqrt{\omega_x^2+\omega_y^2}$ \cite{scith,sci-y}.}  Nevertheless, a supersolid with a highly circularly-asymmetric  distribution of matter $-$ the droplets and the background atom cloud  $-$ has a 
slightly reduced  anomalous moment  of inertia compared to the classical moment of inertia  and a  {reduced}      scissors-mode-oscillation frequency compared to its theoretical estimate.

\begin{figure}[t!]
\begin{center}
\includegraphics[width=.49\linewidth]{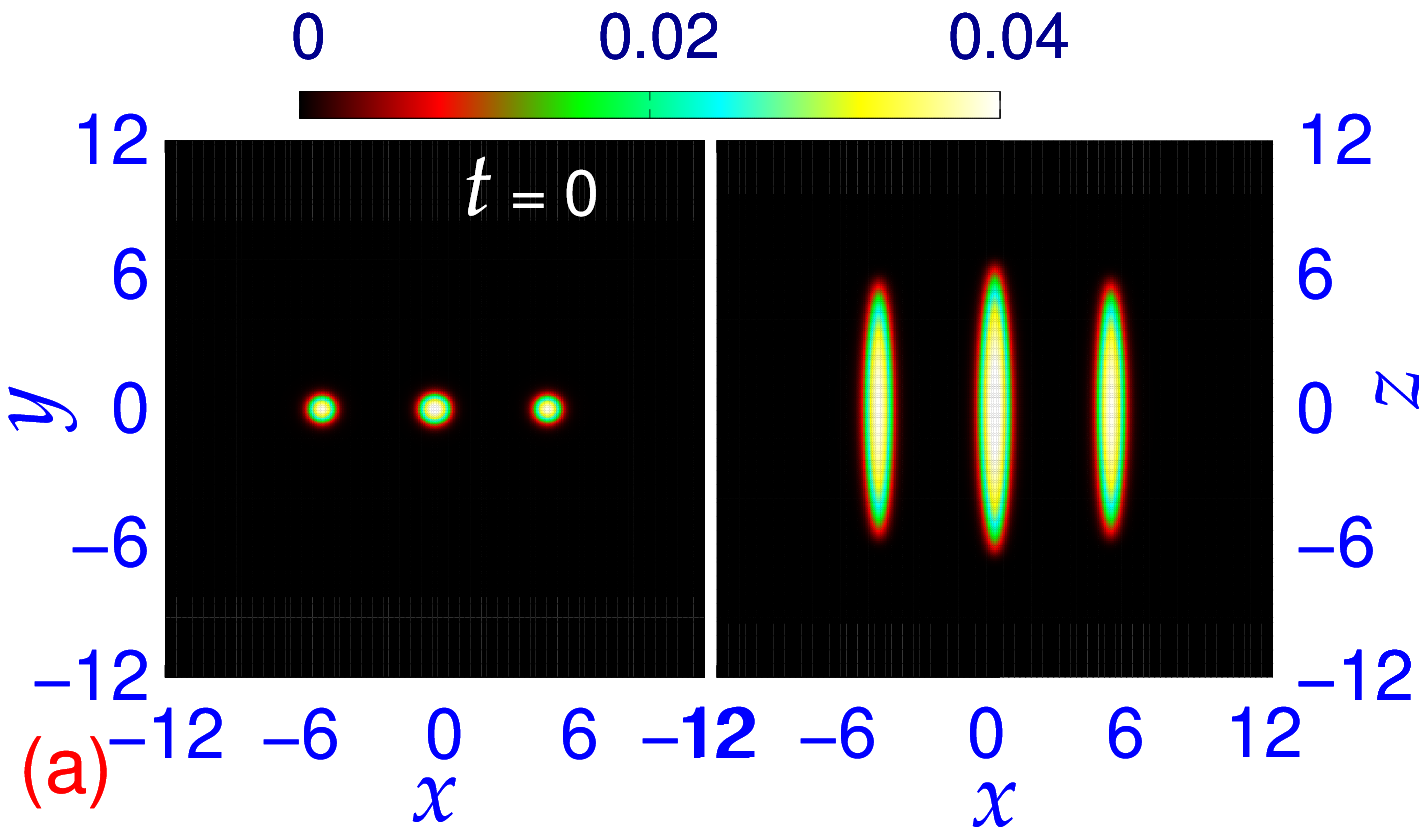}
\includegraphics[width=.49\linewidth]{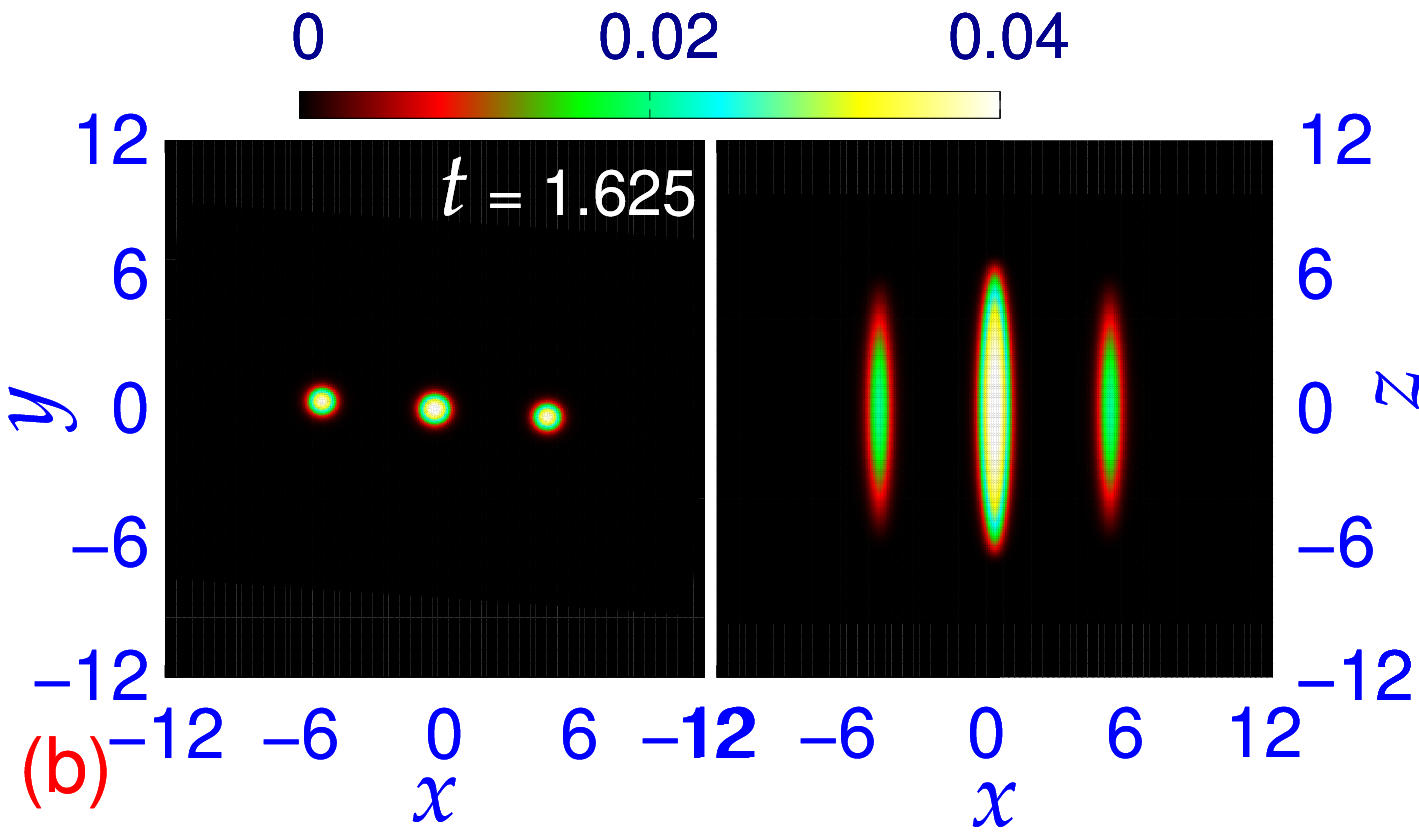}
\includegraphics[width=.49\linewidth]{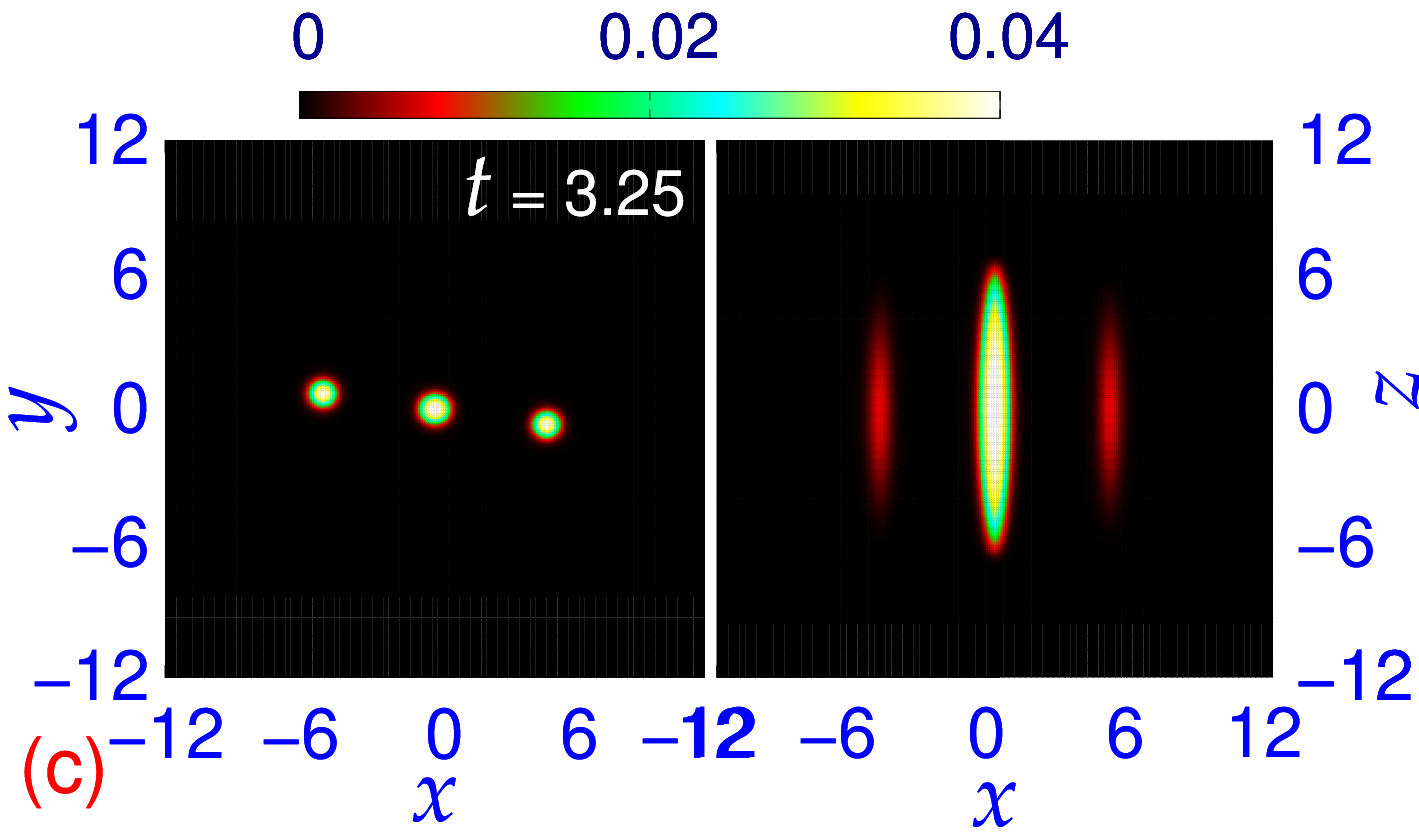}
\includegraphics[width=.49\linewidth]{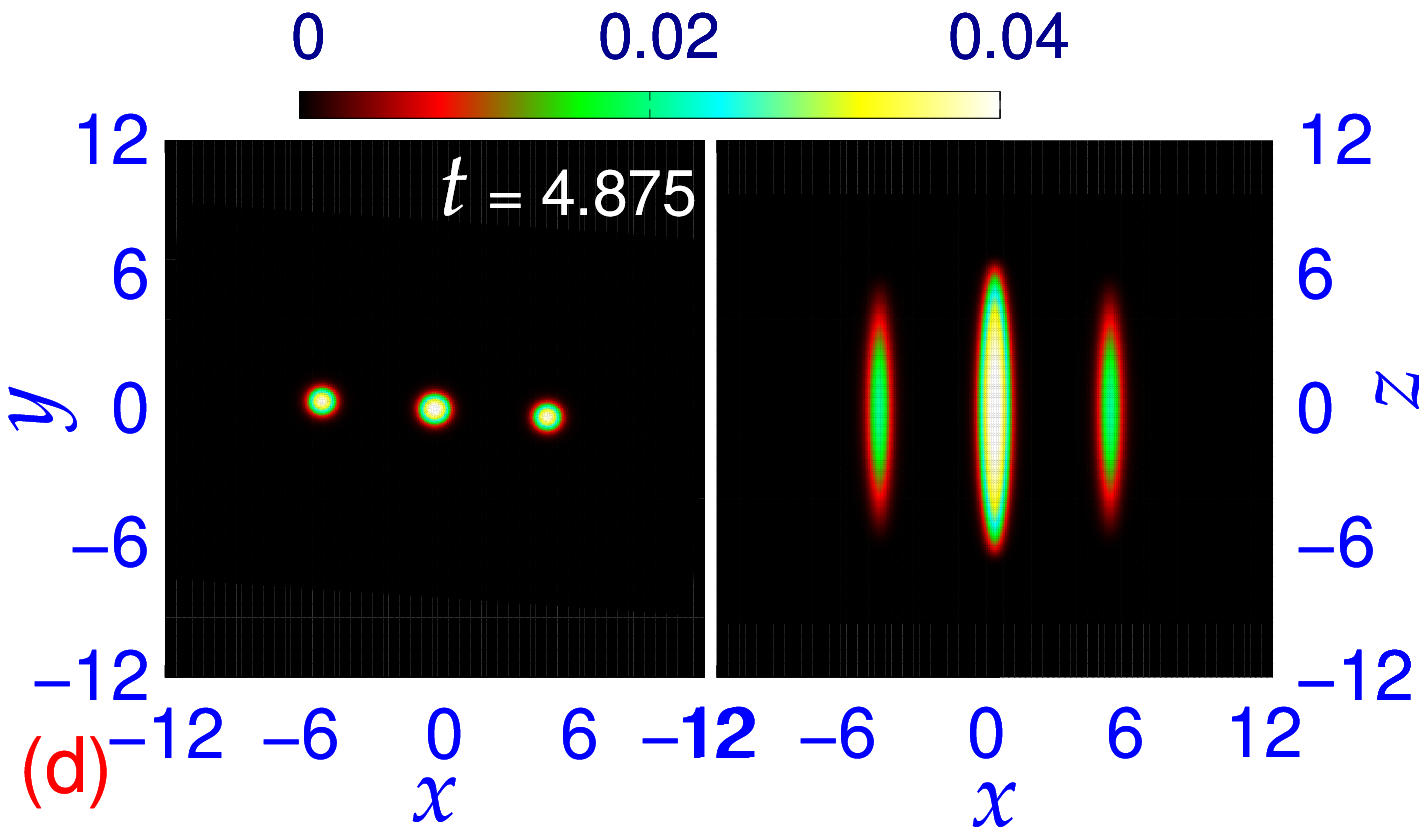}
\includegraphics[width=.49\linewidth]{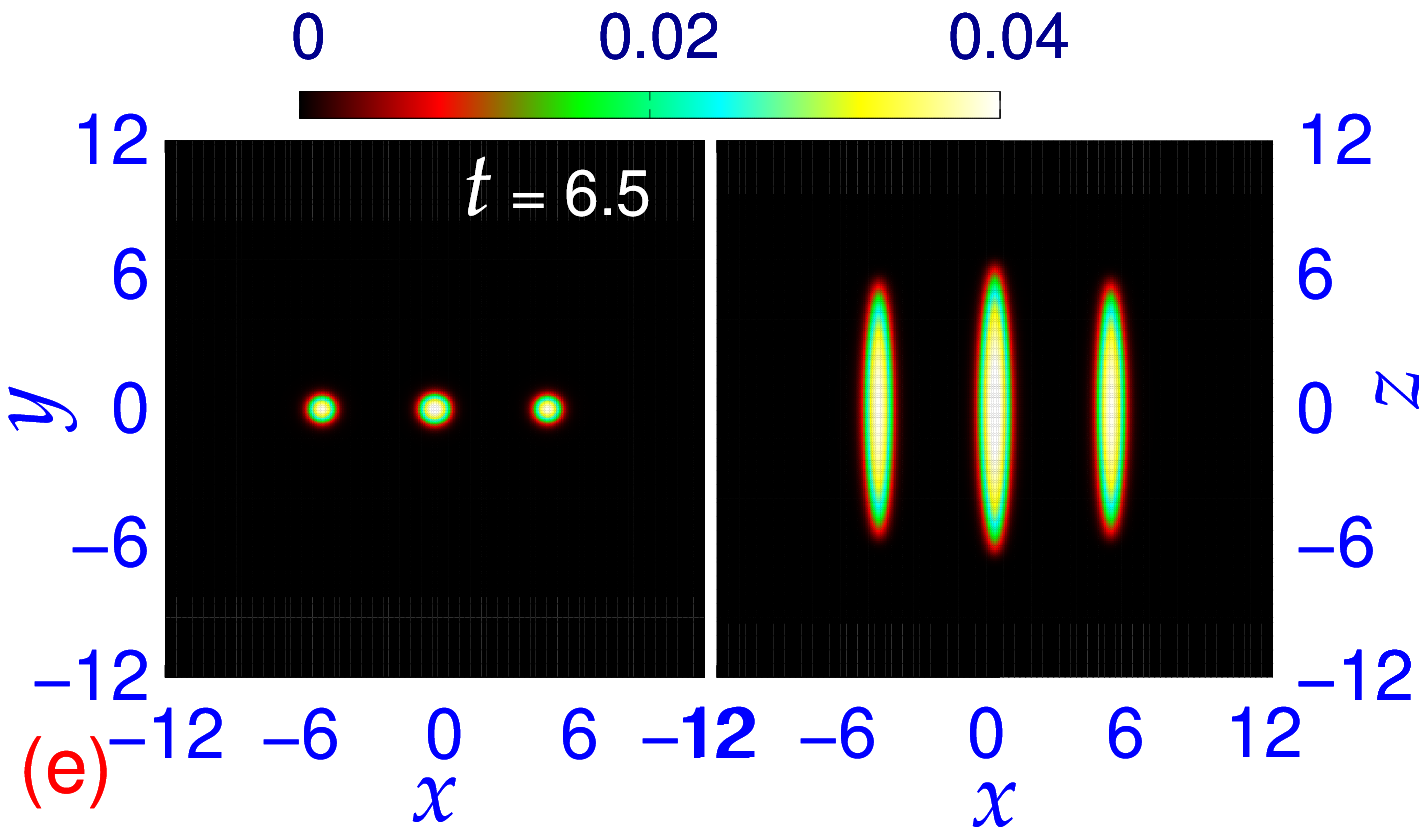}
\includegraphics[width=.49\linewidth]{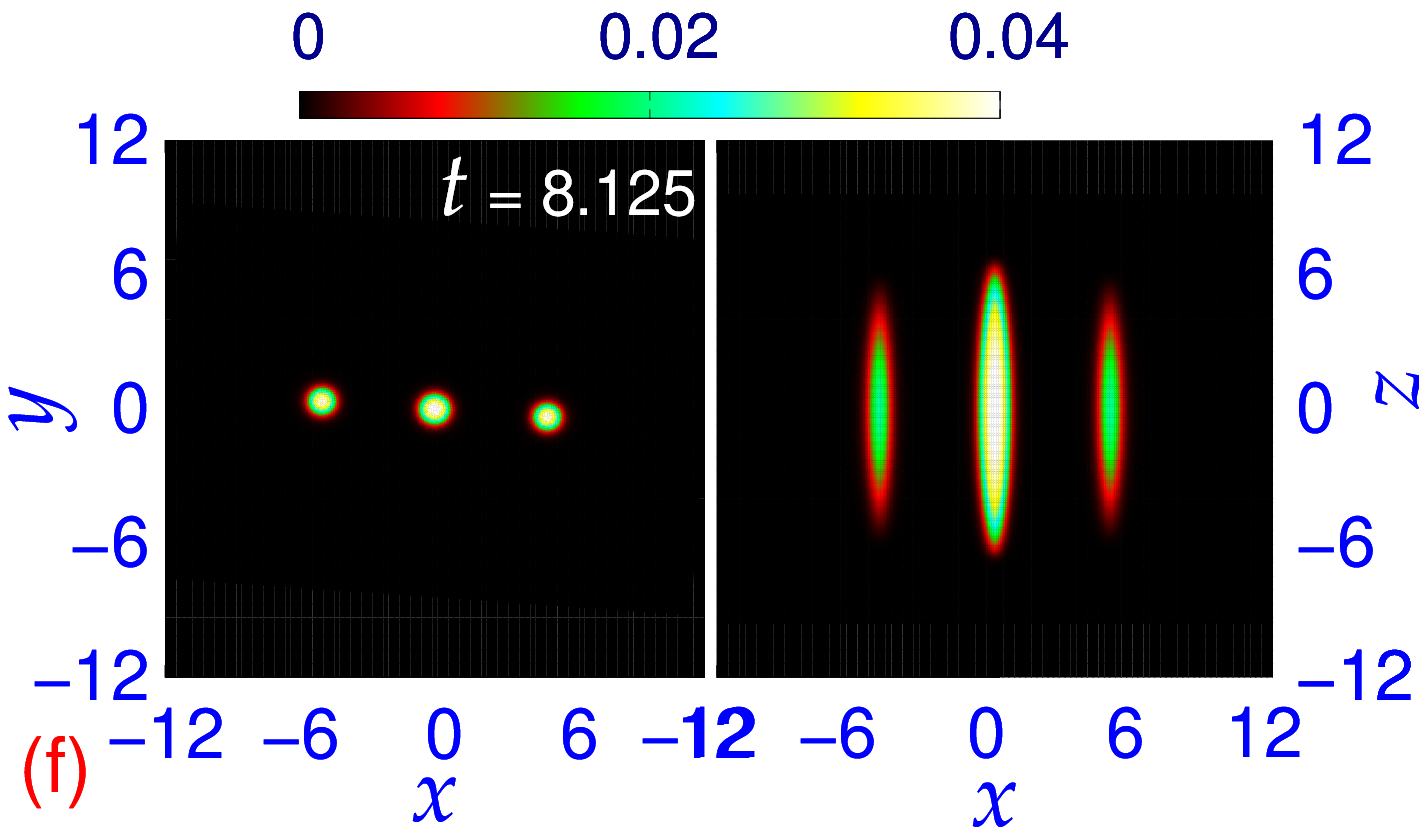} 
 
\caption{ Contour plot of density $|\psi(x,y,0)|^2$ (left side of each {panel})  and $|\psi(x,0,z)|^2$ (right side of each {panel}) 
of the quasi-1D three-droplet supersolid  of $N=35000$ $^{164}$Dy atoms in trap (B) of Fig. \ref{fig5}
executing scissors-mode oscillation
at times (a)
$t=0$, (b) $t=1.625$, (c) $t=3.25$, (d) $t=4.875$, (e) $t=6.5$, (f) $t=8.125$.   { For trap (B) $\omega_x=23/90, \omega_y=\omega_z=1$.}
}
\label{fig6}  
\end{center}
\end{figure}
   
 We consider a quasi-1D three-droplet supersolid in trap (B) and  vary $\omega_y$ in the range 46/90 to 110/90.     
 In Figs. \ref{fig5}(a)-(b) we illustrate  the corresponding  time evolution of
 angle $\theta$ and  energy $E$ for $\omega_x=23/90, \omega_y=\omega_z=1, N=35000$
  fitted to  the periodic oscillation $\cos(0.935\omega_{\mathrm{th}}t)$ and $\cos(2\times 0.935\omega_{\mathrm{th}}t)$, respectively, yielding the frequency $\omega_{\mathrm{sci}}=0.935\omega_{\mathrm{th}}$
 of scissors-mode oscillation with the theoretical frequency $\omega_{\mathrm{th}}= 1.032137899$.  Both the 
 energy and the angle  of the oscillating supersolid are found to execute a steady  sinusoidal oscillation 
as shown in Figs. \ref{fig5}(a)-(b).  The period of angular oscillation  $T=2\pi/\omega_{\mathrm{sci}}= 6.5$ compares well with the theoretical period 
$T_{\mathrm{th}}=2\pi/ \omega_{\mathrm{th}}= 6.0875$.

The  angular oscillation of the supersolid is explicitly
displayed  in  Fig. \ref{fig6} through a contour plot of  densities   $|\psi(x,y,z=0)|^2$ and 
$|\psi(x,y=0,z)|^2$ at times $t=0,1.625,3.25,4.875,6.5,8.125$. The supersolid starts to rotate in the clockwise 
direction at $\theta =0, t=0$, viz. Fig. \ref{fig6}(a), passes through the minimum-energy equilibrium position in the rotated trap 
at $\theta  =-4\degree$ and  $t=1.625$, viz. Fig. \ref{fig6}(b), to the position of maximum angular displacement   of $\theta =-8\degree$ at $t=3.25$, viz. Fig. \ref{fig6}(c).  Then the supersolid turns around and 
again passes through the equilibrium position
of $\theta =-4\degree$ at $t=4.875$, viz. Fig. \ref{fig6}(d), to the initial position of minimum angular displacement   of $\theta =0\degree$ at $t=6.5$, viz. Fig. \ref{fig6}(e) at the end of a complete oscillation. 
The supersolid then turns around again and the
same dynamics is repeated thereafter, viz.   \ref{fig6}(f).  At the position of  maximum angular displacement   of $\theta =-8\degree$ at $t=3.25$   two droplets away from the center of the quasi-1D supersolid 
completely move out of the $x$-$z$ plane and hence only the central droplet is clearly visible in this plane, viz. Fig. \ref{fig6}(c). For a small angular rotation of the trap in the $x$-$y$ plane, this angular oscillation should be simple harmonic \cite{scith}.   The dynamics  of angular displacement $\theta$ and  energy of the quasi-1D supersolid   in Figs. \ref{fig5}(a)-(b) are found to be simple harmonic for the relatively large spatial trap anisotropy ($\omega_x=23/90$ and $\omega_y=1$)  and a moderate angular amplitude of $4\degree$ 
 employed in this study. For a large angle of rotation of the trap $|\theta_0| \gtrapprox 8\degree$ 
 the dynamics ceases to be simple harmonic in nature with a single frequency.

\begin{figure}[t!]
\begin{center}
\includegraphics[width=.49\linewidth]{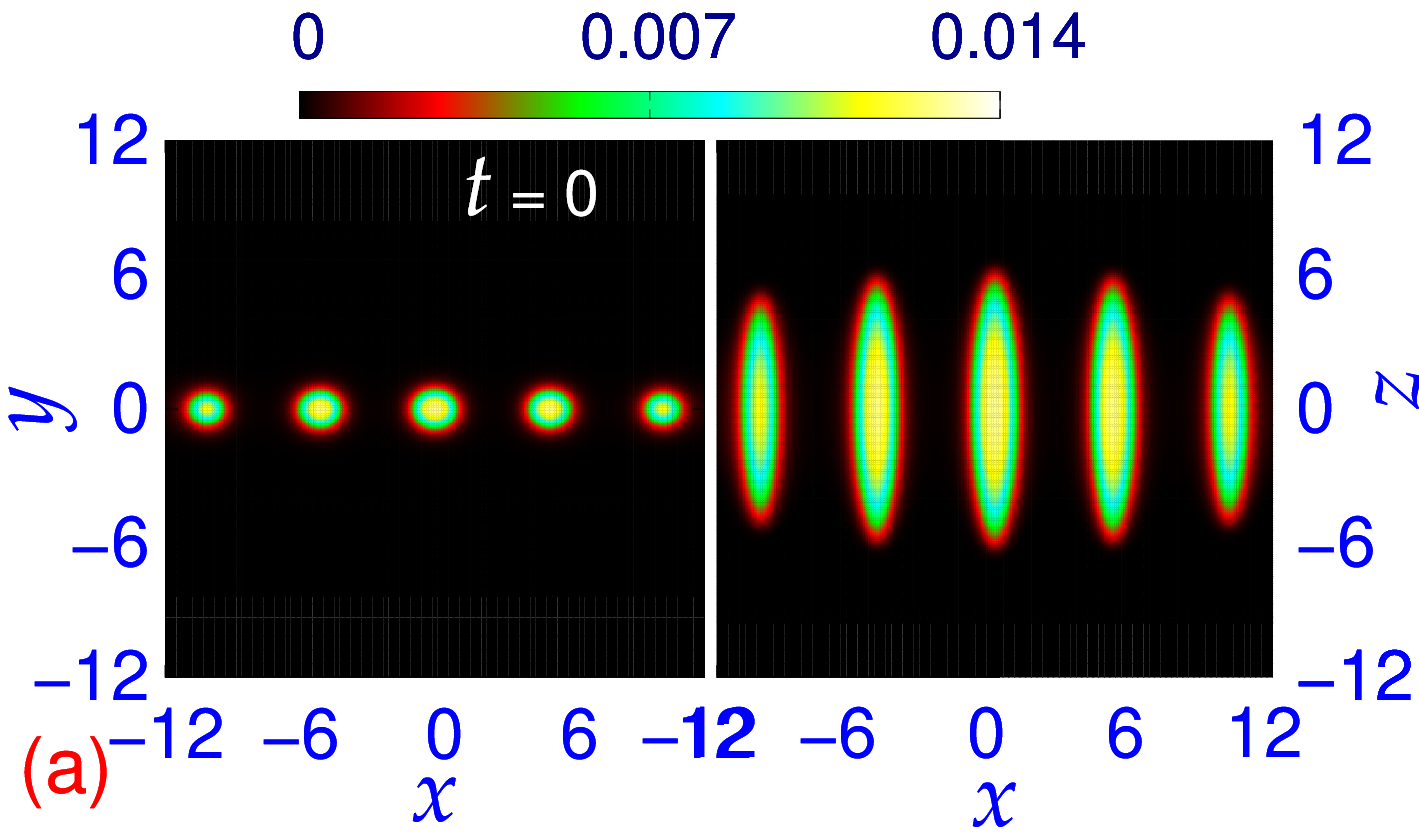}
\includegraphics[width=.49\linewidth]{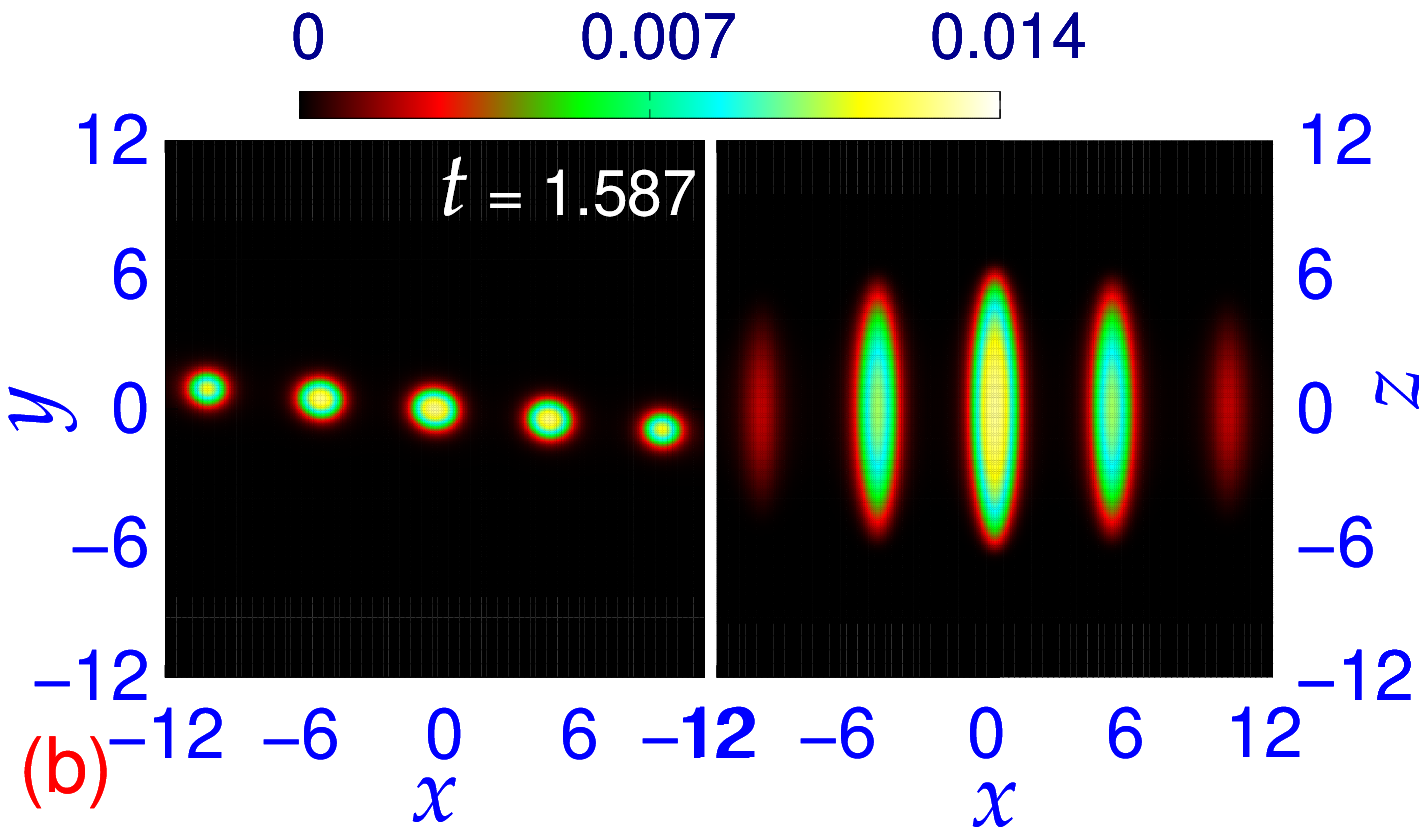}
\includegraphics[width=.49\linewidth]{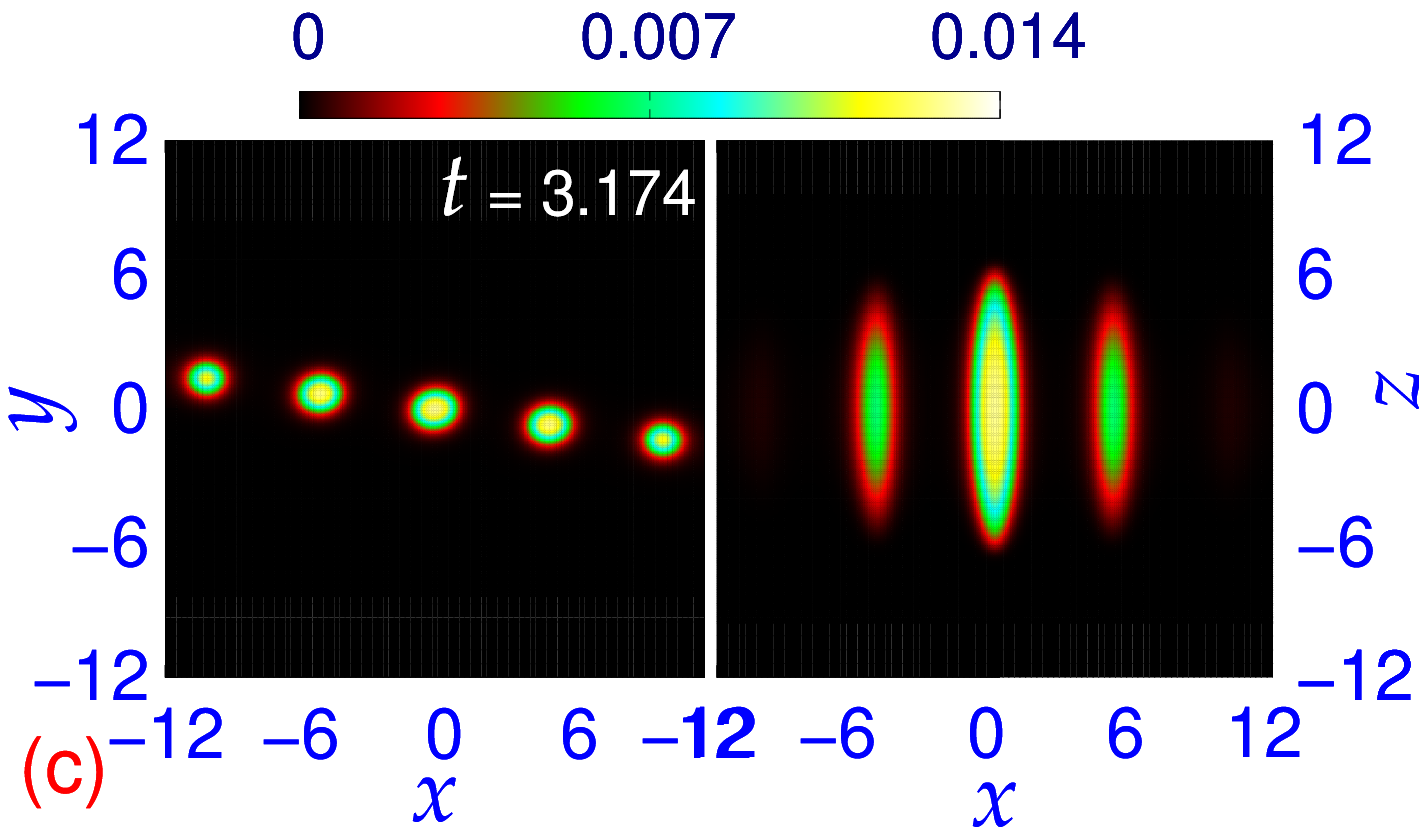}
\includegraphics[width=.49\linewidth]{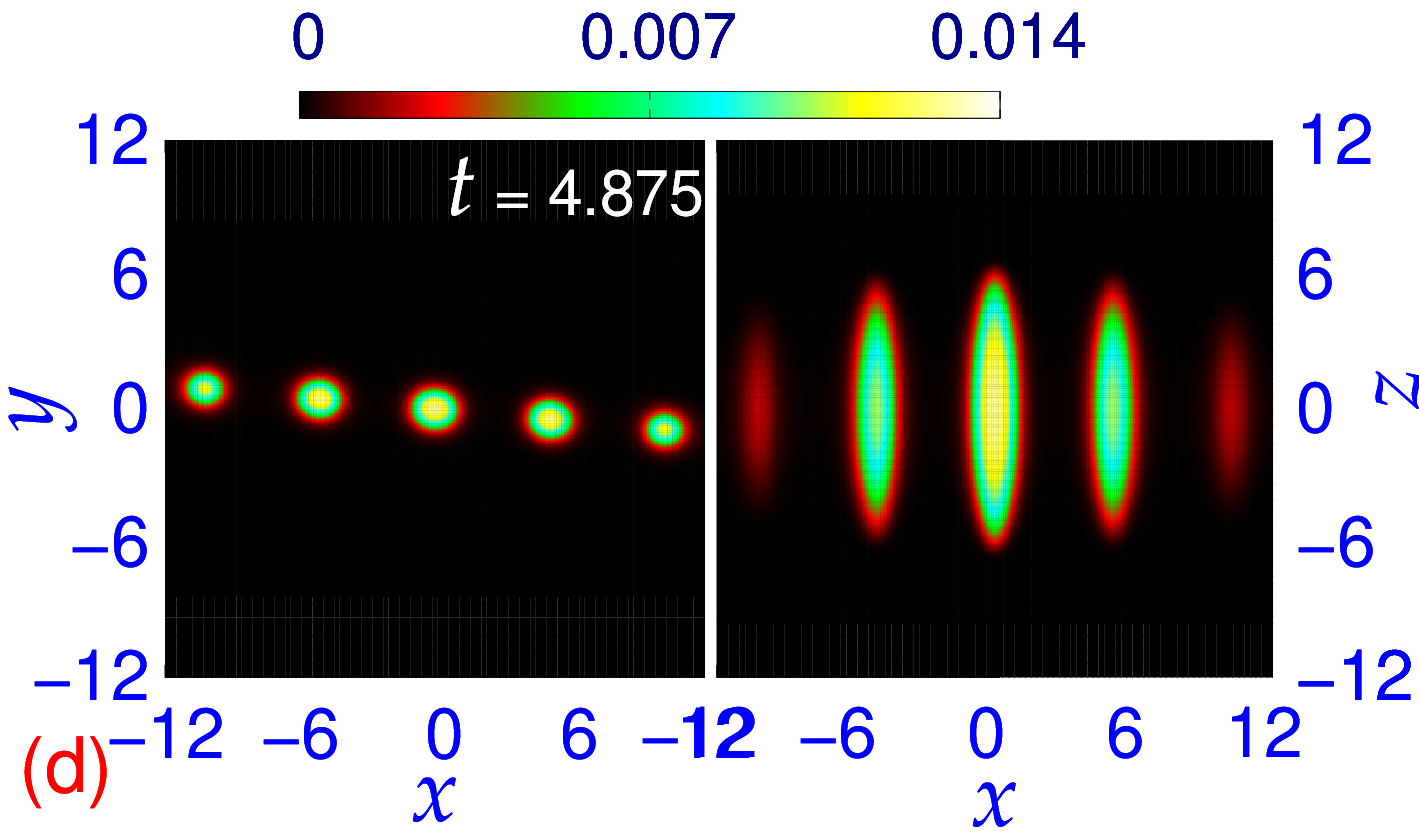}
 \includegraphics[width=.49\linewidth]{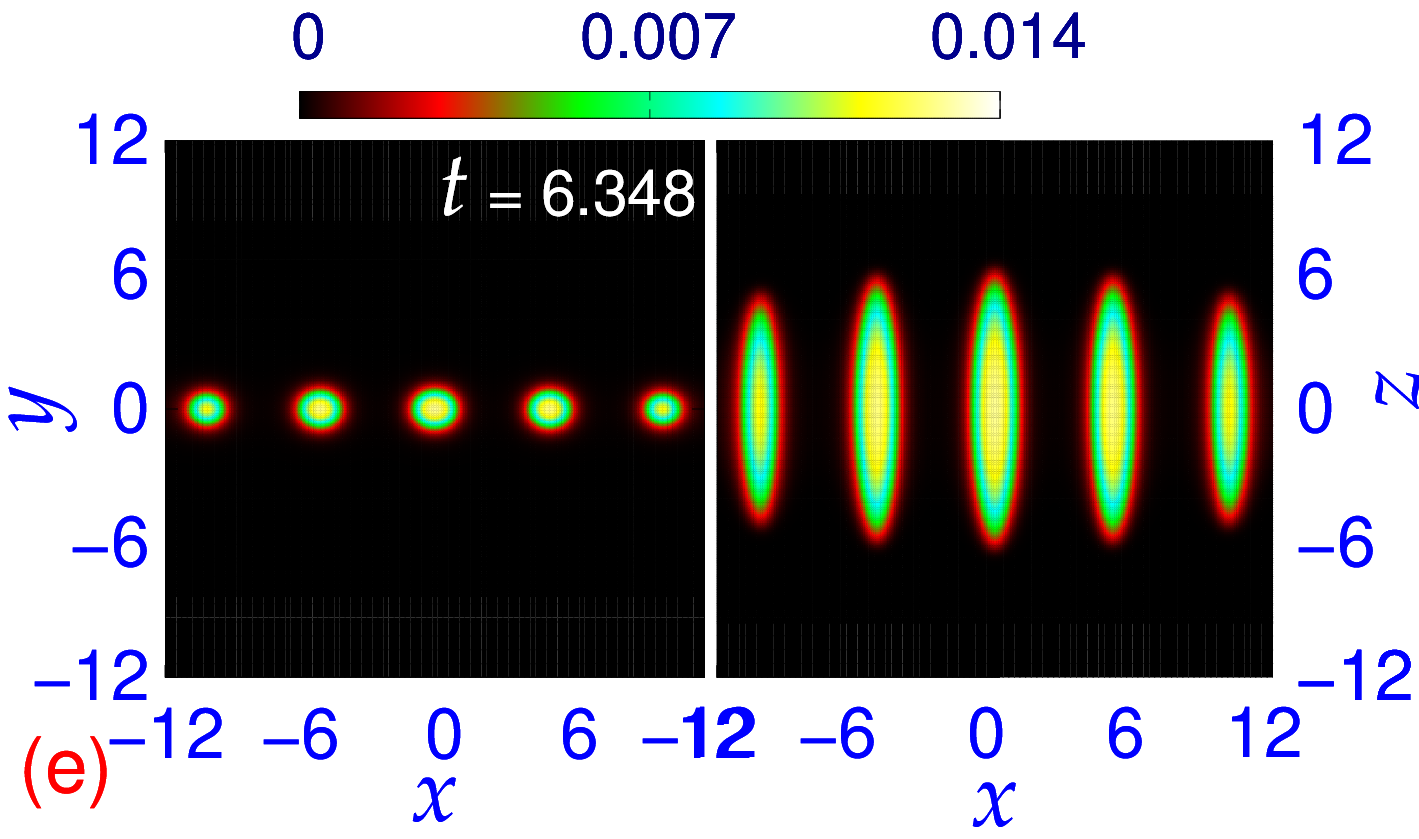}
\includegraphics[width=.49\linewidth]{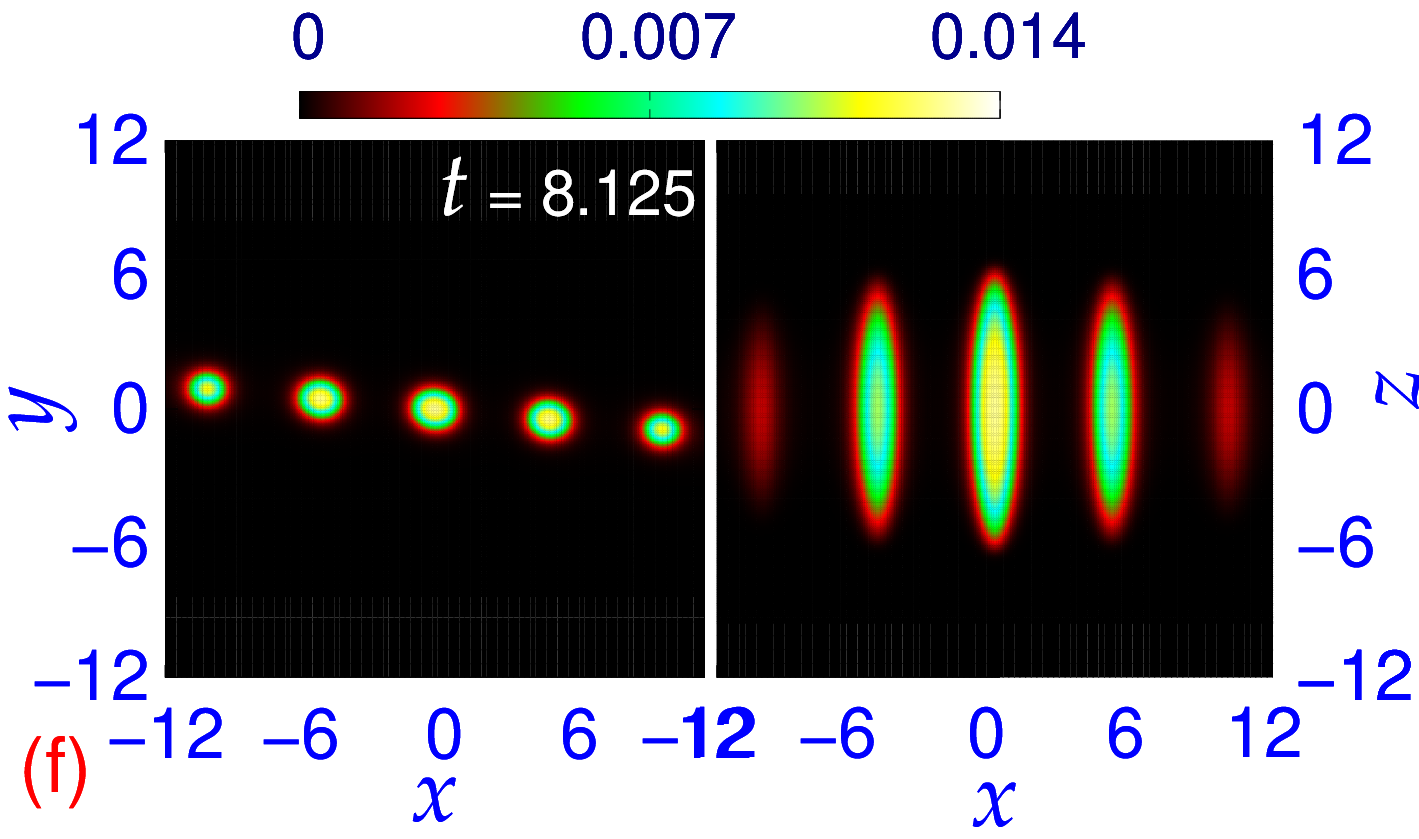}
 
\caption{ Contour plot of density $|\psi(x,y,0)|^2$ (left side of each {panel})   and $|\psi(x,0,z)|^2$ (right side of each {panel}) 
of a quasi-1D five-droplet supersolid   of $N=40000$ $^{164}$Dy atoms of Fig. \ref{fig5} in trap (A) 
executing scissors-mode oscillation
at times (a)
$t=0$, (b) $t=1.625$, (c) $t=3.25$, (d) $t=4.875$, (e) $t=6.5$, (f) $t=8.125$.   {For trap (A) $\omega_x=33/167, \omega_y=\omega_z=1$.}
}
\label{fig8} 
\end{center}
\end{figure}

\begin{figure}[t!]
\begin{center}
\includegraphics[width=.9\linewidth]{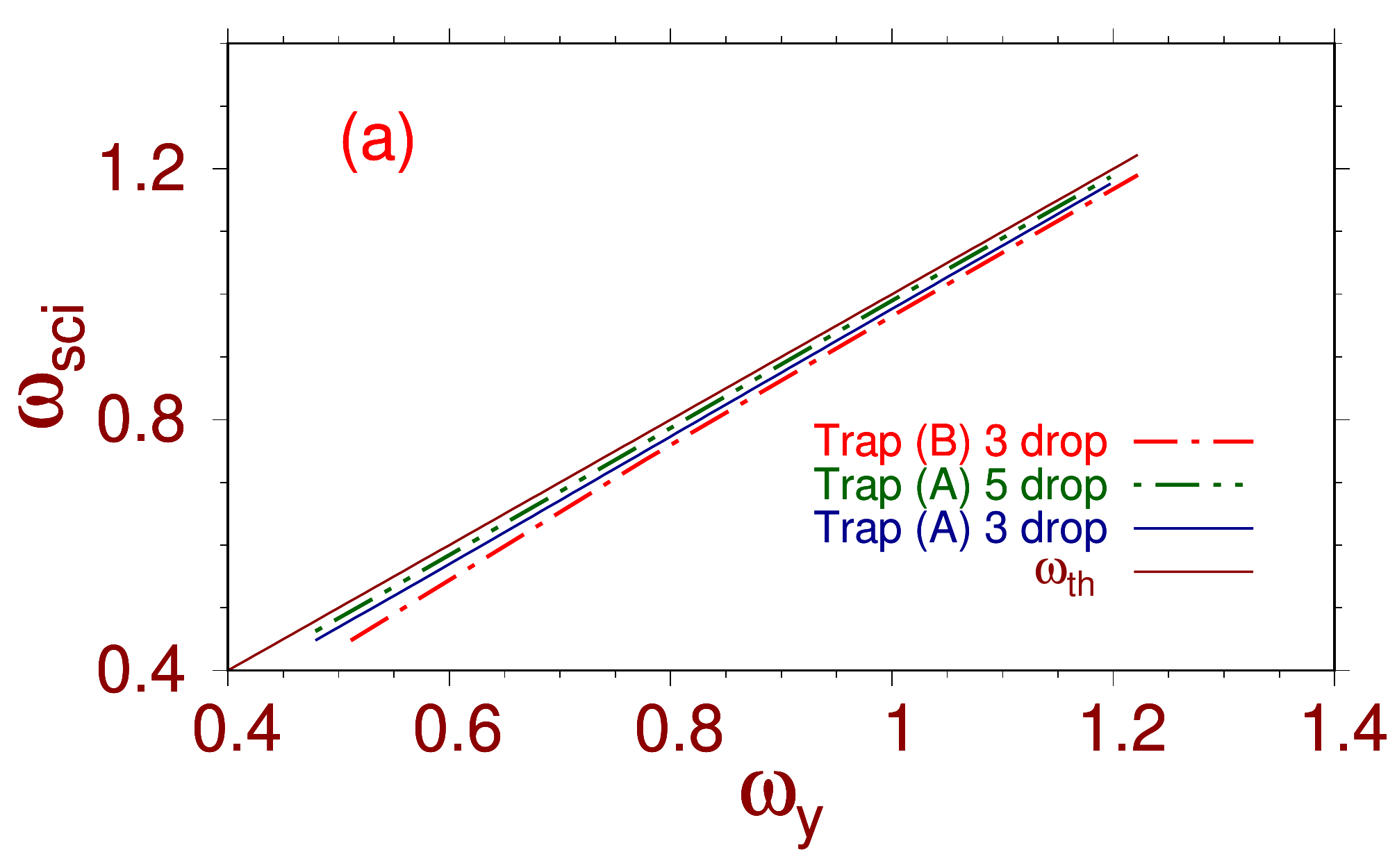}
\includegraphics[width=.9\linewidth]{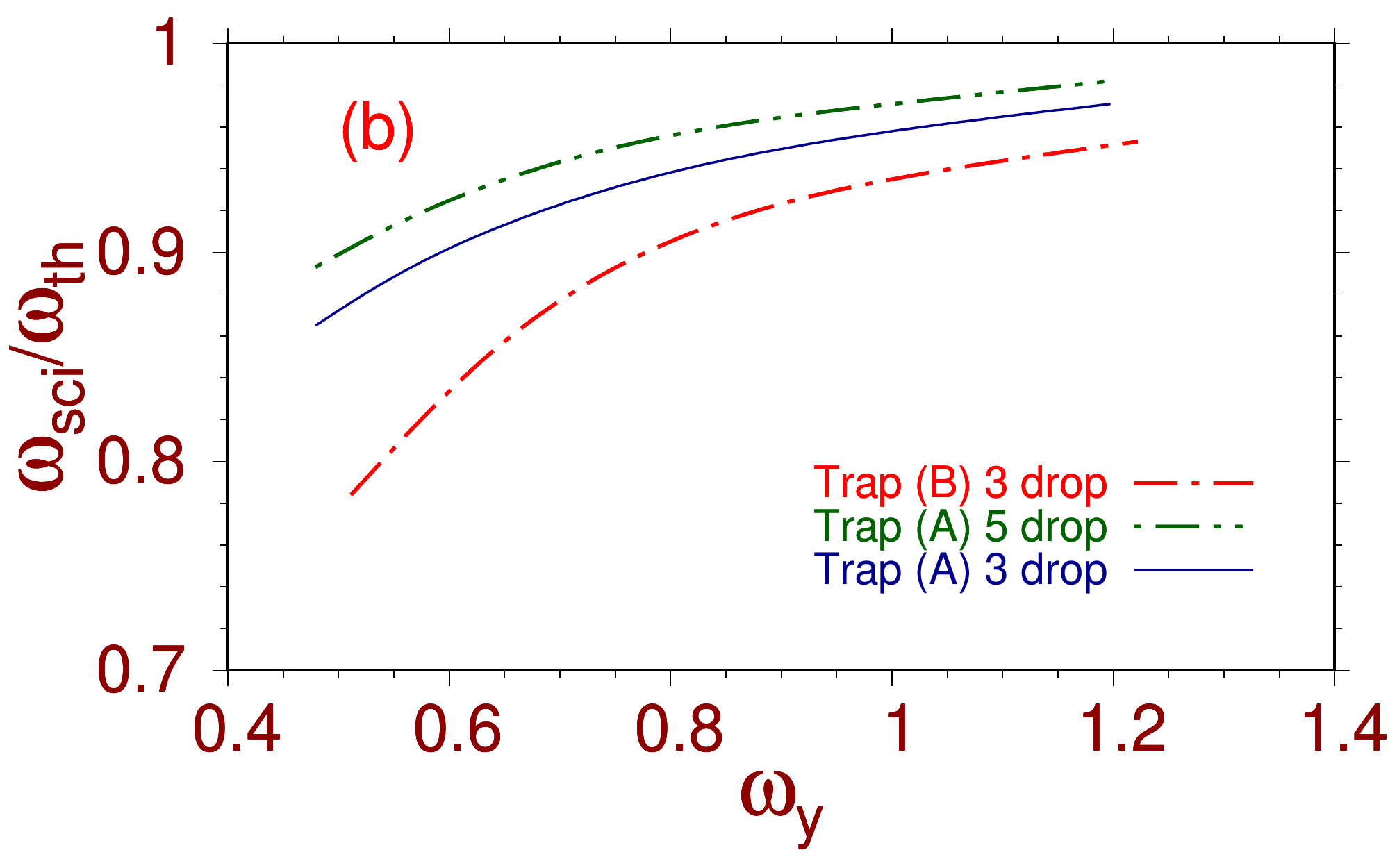}
\caption{ (a) Frequency of scissors-mode oscillation $ \omega_{\mathrm{sci}}$ versus $\omega_y$ for the three-
 and five-droplet states in trap (A) and for the three-droplet state in trap (B), of amplitude $4\degree$,
 compared 
with the theoretical frequency $ \omega_{\mathrm{th}}$. (b) Deviation of scissors-mode oscillation frequency from the theoretical frequency  $ \omega_{\mathrm{sci}} /\omega_{\mathrm{th}}$ versus $\omega_y$ for the three- and five-droplet states in trap (A) and for the three-droplet state in trap (B).  
  In trap (A) $N=25000$ for 3 droplets and 40000 for 5 droplets with $\omega_x=33/167, \omega_z=1, \theta_0 =-4\degree$ and in trap (B) $N=35000$ for 3 droplets with $\omega_x=23/90, \omega_z=1, \theta_0=-4\degree$.}
\label{fig7} 
\end{center}
\end{figure}

 Next we consider the scissors-mode oscillation of a quasi-1D five-droplet supersolid in trap (A) for 
 $\omega_x= 33/167, \omega_y=\omega_z=1, N=40000$  for an initial trap rotation of $\theta_0=-4 \degree$  in detail. (The same for a three-droplet supersolid in trap (A) will not be cosidered here.)   The variation of $\theta$ with time is illustrated in Fig. \ref{fig5}(a) and fitted to a sinusoidal oscillation.  The evolution of this oscillation with time is illustrated by snapshots of contour density plots of $|\psi(x,y,0)|^2$ and 
  $|\psi(x,0,z)|^2$ in Fig. \ref{fig8} at times (a) $t=0$, (b) $t=1.587$, (c) $t=3.174$,  (d) $t=4.875, $
  (e) $t=6.348,$ and (f) $t=8.125$ with angular displacements $\theta =0, \theta \approx -4 \degree, \theta \approx  -8 \degree, \theta \approx -4\degree,\theta  \approx 0$ and $\theta  \approx -4 \degree$, respectively, illustrating the periodic nature of the oscillation.  The period of this oscillation is $T=6.348$,
 compared to the theoretical period of $T=2\pi/\omega_{\mathrm{th}} = 6.1634$,  and the system is back to the initial state $\theta \approx 0$ in Fig. \ref{fig8}(e) at the end of a complete cycle.

\begin{figure}[t!]
\begin{center}
\includegraphics[width=.45\linewidth]{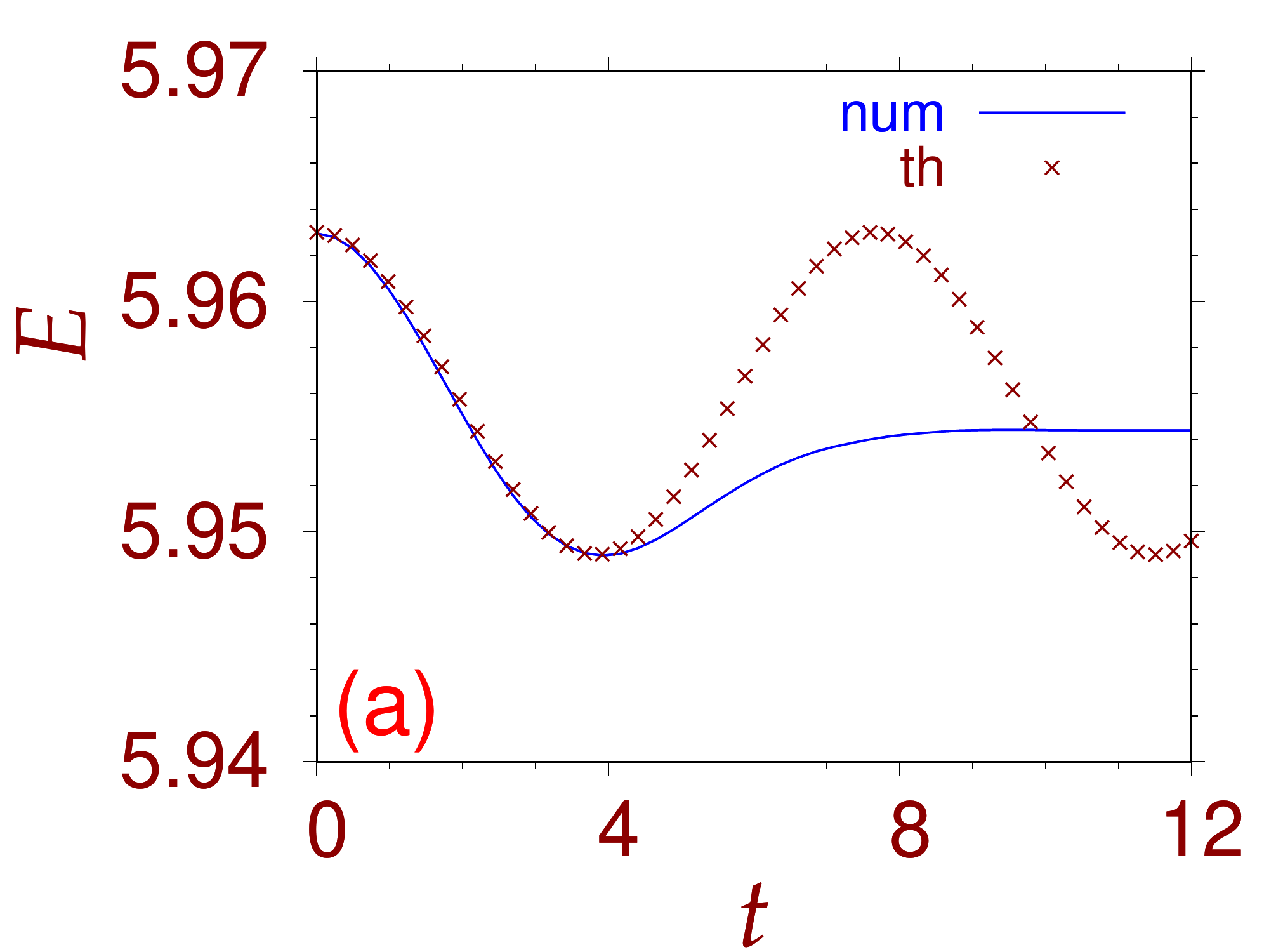}
\includegraphics[width=.32\linewidth]{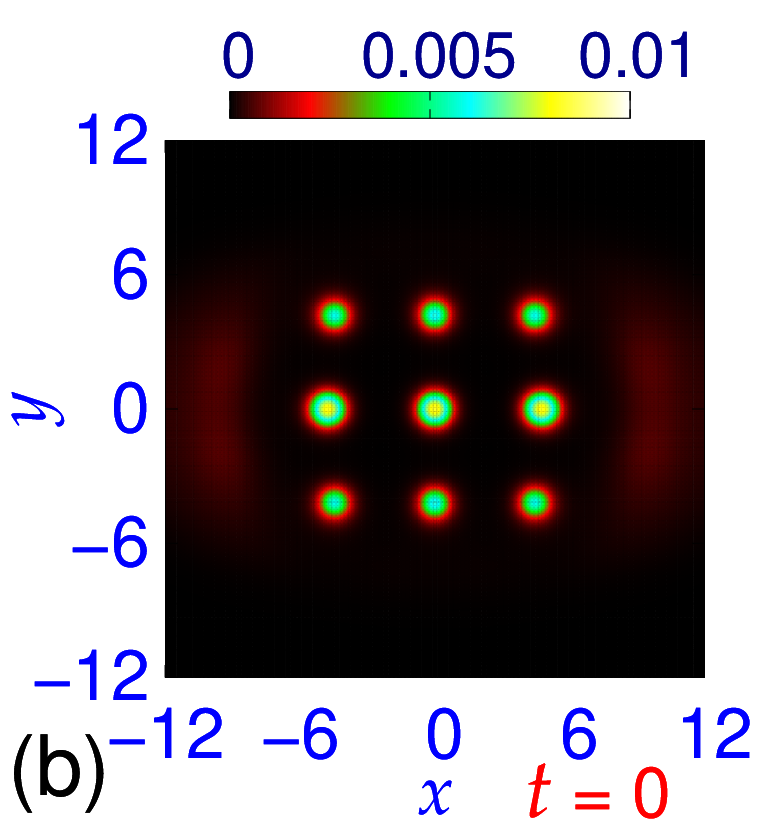}
\includegraphics[width=.32\linewidth]{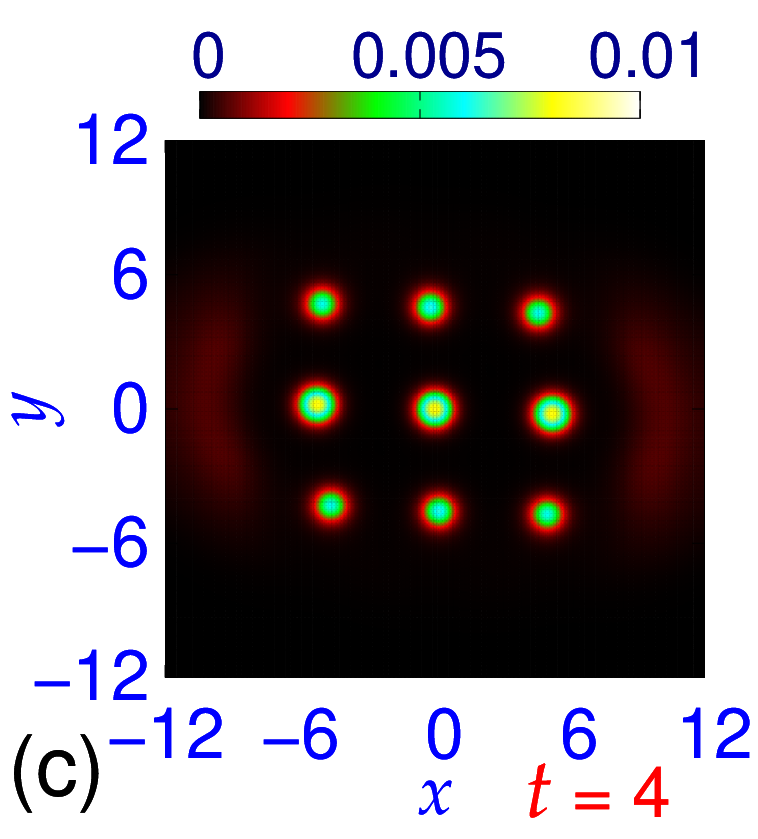}
\includegraphics[width=.32\linewidth]{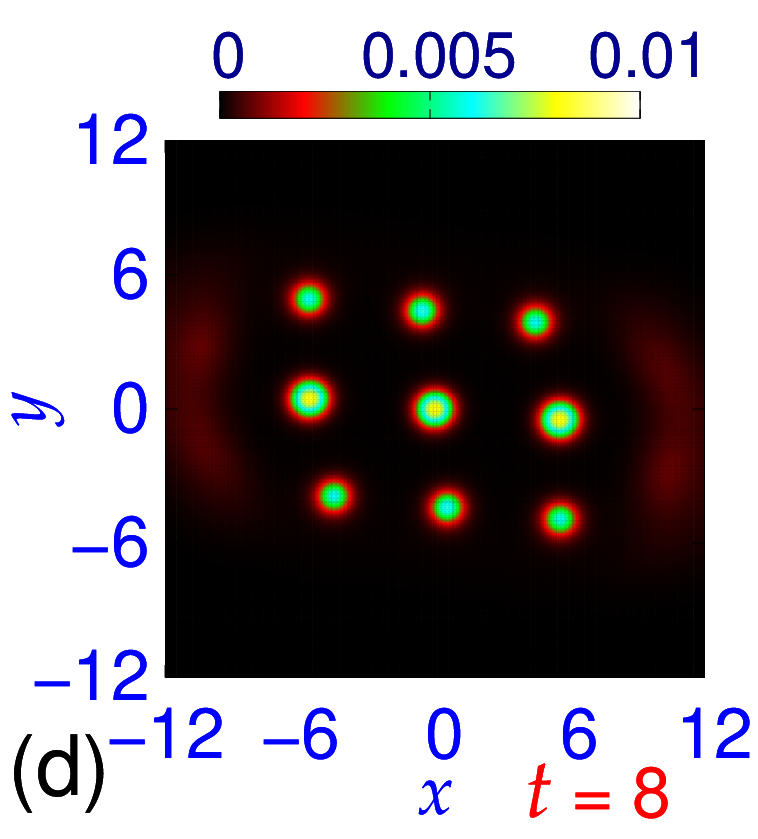}
 \includegraphics[width=.32\linewidth]{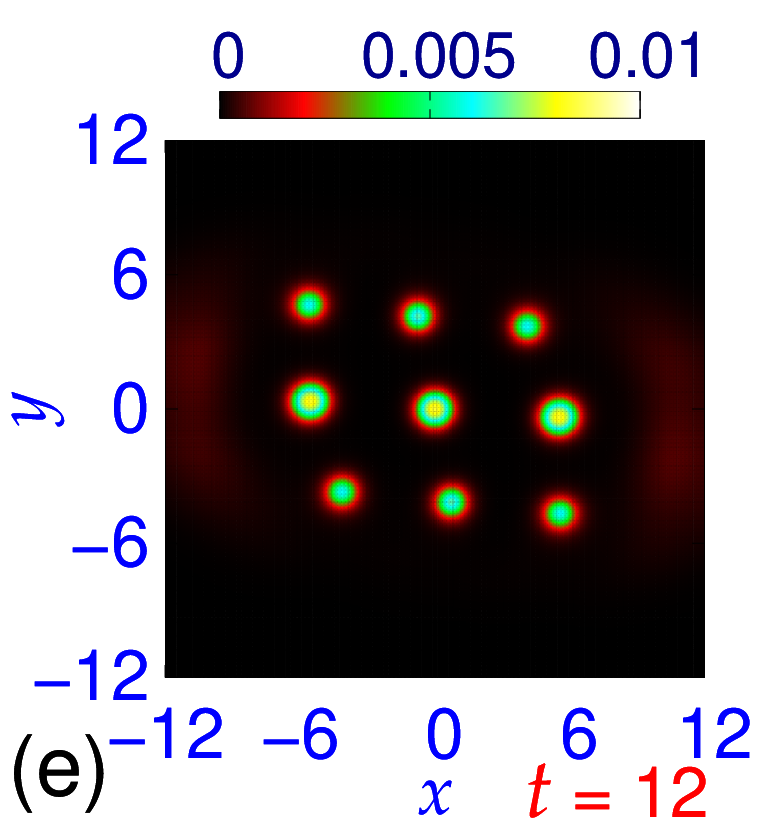}

\caption{ (a) Numerical (num) energy $E$ versus time $t$ during scissors-mode oscillation of a nine-droplet supersolid of $^{164}$Dy atoms of amplitude $4\degree$ in trap (A) fitted to its  theoretical estimate (th) $\cos(2\times \omega_{\mathrm{th}}t)$.
 Contour plot of density                $|\psi(x,y,0)|^2$ 
of this supersolid     
undergoing scissors-mode oscillation 
at times (b)
$t=0$, (c) $t=4$, (d) $t=8$, (e) $t=12$.  The parameters for this simulation are    $\omega_x=33/167, \omega_y=60/167, \omega_z=1$ and $N=60000.$
}
\label{fig9} 
\end{center}
\end{figure}

The variation of the frequency of scissors-mode oscillation $\omega_{\mathrm{sci}}$ of a quasi-1D three-droplet supersolid in trap (A) or trap (B) and of a quasi-1D five-droplet   supersolid in trap (A)
  as a function of  frequency $\omega_y$  is illustrated in Fig. \ref{fig7}(a) and is compared with the theoretical frequency $\omega_{\mathrm{th}}$. The deviation of the scissors-mode oscillation frequency $\omega_{\mathrm{sci}}$
from the theoretical frequency  $\omega_{\mathrm{th}}$, e.g. $\omega_{\mathrm{sci}}/\omega_{\mathrm{th}}$, for different $\omega_y$ is presented in Fig. \ref{fig7}(b). {The actual scissors-mode frequency
is always less \cite{string} than its  theoretical estimate $\omega_{\mathrm{th}}> \omega_{\mathrm{sci}}$}, as can be found in Fig. \ref{fig7}. 
  The theoretical frequency $\omega_{\mathrm{th}}$ in both traps leads essentially to the same line shown in Fig. \ref{fig7}(a). Although, the theoretical frequency $\omega_{\mathrm{th}}$ is a good approximation to the actual frequency  $\omega_{\mathrm{sci}}$ for a wide range of variation of trap parameters, 
 as can be seen in Fig. \ref{fig7}(a), 
 the agreement improves,  as $\omega_y$ increases,  resulting in an  increase of the  asymmetry of the trap in the  $x$-$y$ plane.  For the same trap,  viz. trap (A),
 the agreement improves  as  the number of droplets   increases  resulting in a larger supersolid.  For the same frequency $\omega_y$, the agreement also improves  in a stronger trap, e.g., trap (A) with an overall 
 stronger trapping (large $\sqrt[3]{\omega_x\omega_y\omega_z}$) compared to trap (B). 
 Nevertheless, for the same number of droplets, the frequency of scissors-mode oscillation was found to be  independent of the number of atoms. { The trap (B) is the same as in Ref. \cite{sci-y} and it is possible to compare the present results obtained in  trap (B)  with that reference. In trap (B), with $\omega_x=23/90= 0.2556, \omega_y =46/90=0.5111, \omega_z=1,$ and  $a= 87.2a_0$ ($\varepsilon_{\mathrm{dd}}=1.5$)  they obtained in their experiment  $\omega_{\mathrm{sci}}/\omega_{\mathrm{th}}\approx 0.78 \pm 0.03 $.  
  For  the same trap with   $a= 85a_0$  we get $\omega_{\mathrm{sci}}/\omega_{\mathrm{th}}\approx 0.784 $.  We also repeated our calculation in the same trap  for  $a=87.2a_0$ as in Ref. \cite{sci-y} and we found  $\omega_{\mathrm{sci}}/\omega_{\mathrm{th}}= 0.795 $  in good agreement with their result.   
 We also compared the present results with Ref. \cite{string} for $a=85a_0$ where they used the trap frequencies $\omega_x=20/80=0.25, \omega_y =40/80=0.5,  \omega_z = 1$ pretty close to the present  frequencies for trap (B) in dimensionless units.  Their result of $\omega_{\mathrm{sci}}/\omega_{x}\approx 1.74 $  \cite{string}  for $a=85a_0$ translates to  $\omega_{\mathrm{sci}}/\omega_{\mathrm{th}}  \equiv  \omega_{\mathrm{sci}}/\sqrt{\omega_x^2+\omega_y^2} 
 \approx 0.778 $ in good agreement with the present result $\omega_{\mathrm{sci}}/\omega_{\mathrm{th}}\approx 0.784. $  } 
  In Fig. \ref{fig7} $\omega_y \gtrapprox  0.5$ typically represents a quasi-1D trap, 
  and  it 
  is also the region of sustained periodic scissors-mode oscillation.    
  As   $\omega_y $ decreases 
below a lower limit or increases above an upper limit  
the scissors-mode oscillation ceases to be simple harmonic with a single frequency and becomes irregular in nature. 
The domain of frequencies $0.4 \lessapprox \omega_y \lessapprox 1.2$  seems to be appropriate for a sustained   periodic scissors-mode oscillation.  
  A quasi-2D supersolid can be formed for $\omega_y \lessapprox 0.4 $, while the trap approximates  a quasi-2D 
 configuration; however, in this region   one cannot have a sustained  scissors-mode oscillation as the trap anisotropy in the $x$-$y$ plane is smaller that required for a sustained oscillation. This is why we  could not find any scissors-mode oscillation for a quasi-2D supersolid with square or hexagonal
symmetry.

\begin{figure}[t!]
\begin{center}
\includegraphics[width=.45\linewidth]{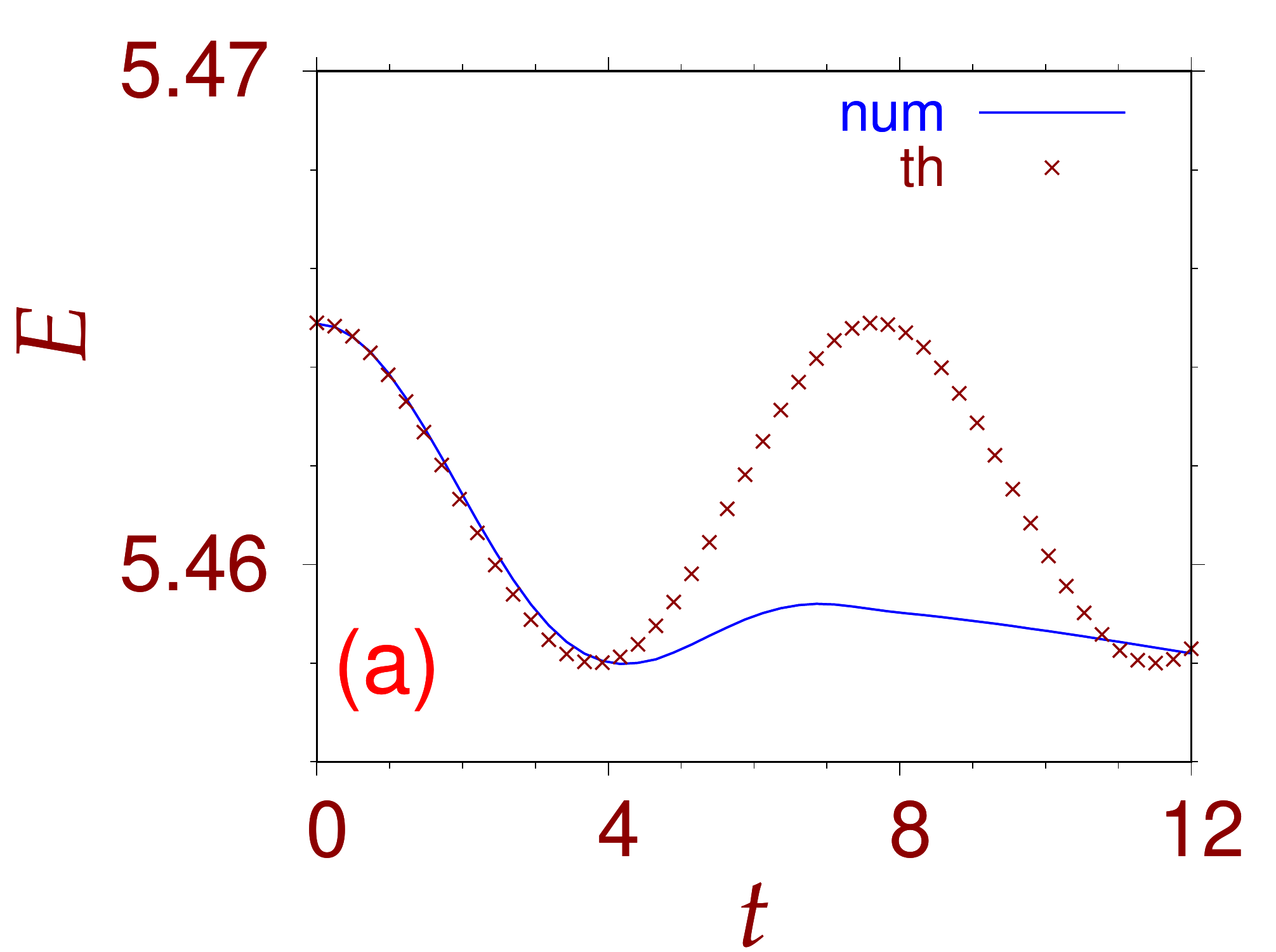}
\includegraphics[width=.32\linewidth]{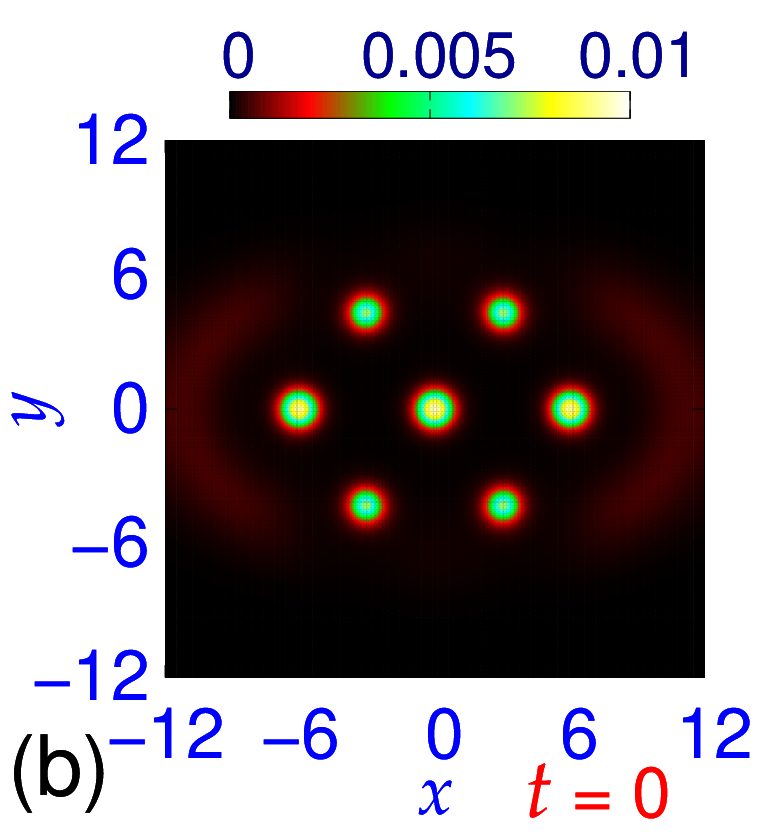}
\includegraphics[width=.32\linewidth]{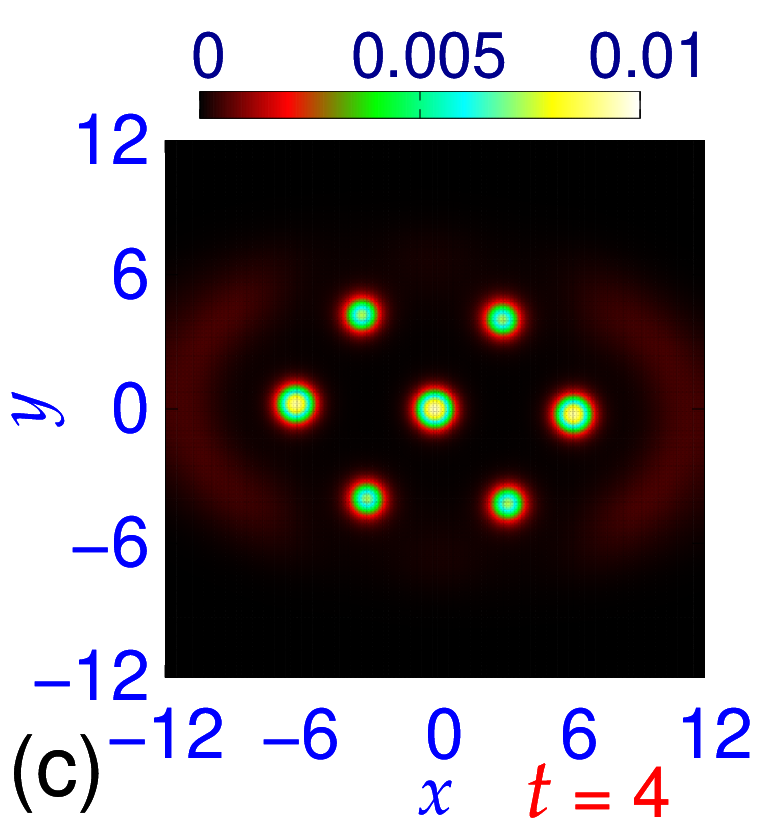}
\includegraphics[width=.32\linewidth]{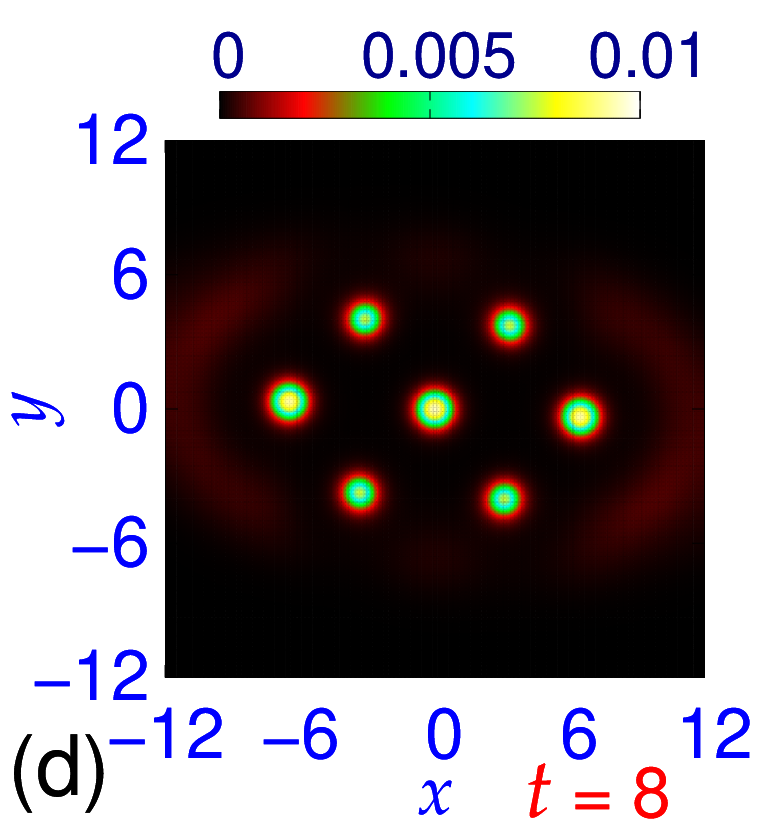}
 \includegraphics[width=.32\linewidth]{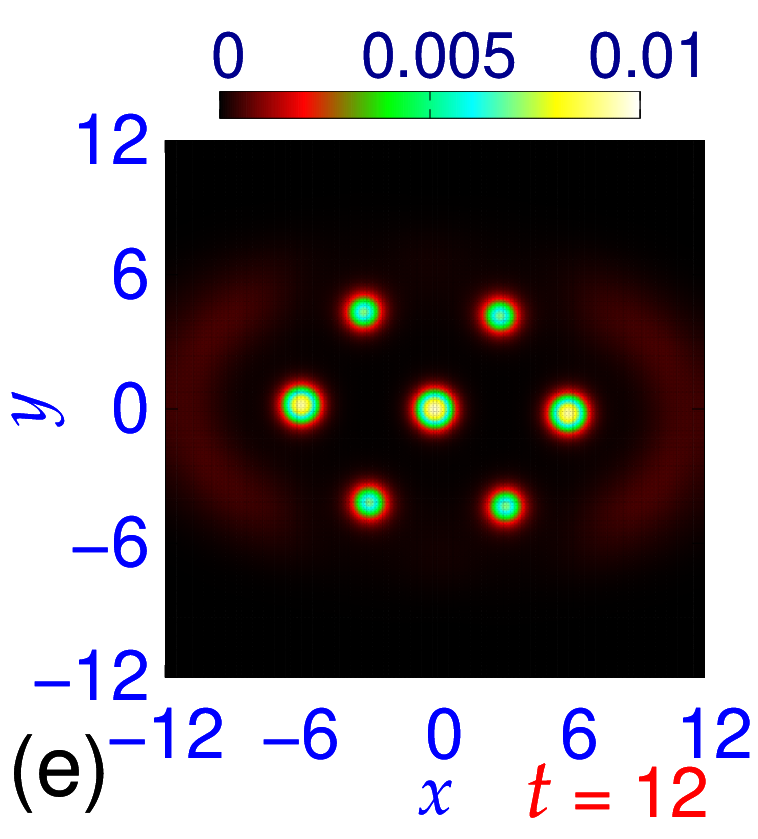}

\caption{ (a) Numerical (num) energy $E$ versus time $t$ during scissors-mode oscillation of a seven-droplet supersolid of  $N=50000$  $^{164}$Dy atoms of amplitude $4\degree$ in trap (A), with $\omega_y=60/167$, fitted to its  theoretical estimate (th) $\cos(2\times \omega_{\mathrm{th}}t)$.
 Contour plot of density                $|\psi(x,y,0)|^2$ 
of this supersolid     
undergoing scissors-mode oscillation 
at times (b)
$t=0$, (c) $t=4$, (d) $t=8$, (e) $t=12$.  
}
\label{fig10} 
\end{center}
\end{figure}

     To illustrate the breakdown of scissors-mode oscillation for a quasi-2D supersolid 
       explicitly, we consider a quasi-2D nine-droplet  supersolid for $\omega_x=33/167, \omega_y=60/167\approx 0.35928, \omega_z=1$ and $N=60000.$   Although a smaller  
  $\omega_y$ is favored for the formation of a quasi-2D nine-droplet supersolid, no regular simple-harmonic scissors-mode oscillation could be found there. { In this case, for a trap-rotation angle of  $\theta_0=-4\degree$, the evolution of energy is compared to its theoretical estimate $\cos(2\omega_{\mathrm{th}}t)$  in  Fig. \ref{fig9}(a), where there is no periodic oscillation   indicating  a breakdown of the scissors-mode oscillation,  which can be seen more explicitly from the contour plot of density $|\psi(x,y,0)|^2$ at times (b) $t=0$, (c) $t=4$, (d) $t=8$, and (e) $t=12$. }
 Due to a stronger trap in the $y$ direction, the $y=0$ droplets are larger, accommodating a larger number of atoms than the $y\ne 0$ droplets, as can be seen in the initial density in Fig. \ref{fig9}(b). 
 In Fig. \ref{fig9}(a) we see  that the oscillation agrees with the theoretical estimate upto $1/4$ cycle  and after this  the oscillation becomes irregular.
 At $t=4$ in (c), at the end of $1/4$ cycle, the supersolid rotated about $4\degree$. After that the rotation angle increases to about $6\degree$ at $t=8$ in (d) and remains roughly the same thereafter, viz. (e) at $t=12$,  indicating a breakdown of the scissors-mode oscillation.

We also studied the scissors-mode oscillation of a seven-droplet triangular-lattice supersolid 
state  in  trap (A) for $N=50000$ $^{164}$Dy atoms.  In this case the panorama is, quite similar to 
the case  studied in Fig. \ref{fig9},  as exhibited in Fig. \ref{fig10}
through   (a) the evolution of  energy and contour plot of density $|\psi(x,y,0)|^2$ at times (b) $t=0$,
(c) $t=4$, (d) $t=8$, and (e) $t=12$.   Again a  lack of periodic oscillation in energy in Fig. \ref{fig10}(a) indicates a breakdown of the scissors-mode oscillation.  The oscillation is regular upto $t=4$, viz. Fig. \ref{fig10}(c), where the angular displacement of the supersolid is $4\degree$. After that the angular displacement increases to about $5\degree$ and the oscillation practically stops there, viz. Figs. \ref{fig10}(d)-(e),  as in the case of the nine-droplet supersolid in Fig. \ref{fig9}. { Similar results were also found  in  related studies \cite{related,related2}.}
 The breakdown of scissors-mode oscillation in these cases of a quasi-2D dipolar supersolid does not mean a lack of superfluidity, because the confining trap does not have a large asymmetry in the   $x$-$y$ plane as required for a sustained scissors-mode oscillation.

\section{Summary}

\label{IV}

In this paper we studied the linear  {dipole-mode} and angular scissors-mode oscillation dynamics of a  dipolar supersolid using a  beyond-mean-field model including the LHY interaction with a view to test both the superfluidity and the rigid solid structure of the material. The LHY interaction  has a higher-order quartic nonlinear term compared to the cubic nonlinear term of the mean-field model. The quartic nonlinearity leads to a higher-order short-range repulsion that stops the collapse  instability resulting from the strong dipolar attraction in presence of a moderate short-range repulsion resulting from the cubic nonlinear term. In the case of  {dipole-mode} oscillation,
both the quasi-1D and quasi-2D supersolids passed the above-mentioned test  with honors demonstrating the superfluidity and  rigid solid structure of the supersolid. Only the  quasi-1D dipolar supersolid  was capable of executing a stable angular scissors-mode oscillation  without any deformation of the crystalline structure
in a highly asymmetric trap in the $x$-$y$ plane.  The relatively large asymmetry of the trap is also necessary for a sustained angular scissors-mode oscillation of a nondipolar BEC. However, in such a trap, it is not possible to have a quasi-2D supersolid; a quasi-2D supersolid naturally appears in a  
 trap with small asymmetry in the $x$-$y$ plane, but such a trap is found to be not appropriate for a sustained  angular scissors-mode oscillation. 

In the study of linear  {dipole-mode} oscillation,  we considered a quasi-1D three-droplet supersolid and a quasi-2D nine-droplet supersolid in trap (A) \cite{2d3,drop1}. The study of angular scissors-mode oscillation was performed with a quasi-1D three- as well as a five-droplet supersolid  in trap (A) and  with a quasi-1D three-droplet supersolid in trap (B) with appropriate parameters. The  {dipole-mode} oscillation was started 
by giving an initial displacement of $x_0=5$  of  the trap along the $x$ direction perpendicular to the polarization direction $z$. 
The angular scissors-mode oscillation was started with an initial rotation of the  trap by $\theta_0=-4\degree$ around the $z$ axis. For a periodic scissors-mode oscillation, the trap should be asymmetric in the $x$-$y$ plane 
with a moderate anisotropy between an upper and lower limits. 
The frequency of  the  {dipole-mode} oscillation was found to be identical with the theoretical frequency $\omega_x$ in all cases; that of  the scissors-mode oscillation was close  to, but always less than, { its theoretical estimate $\omega_{\mathrm{th}},$  as also found in other studies \cite{sci-y,string}.}  
 The  deviation of the scissors-mode frequency from its theoretical estimate reduces as the asymmetry of the trap in the $x$-$y$ plane increases  and also  as the overall strength of the trap increases from trap (B) to trap (A). {The frequency of the scissors-mode oscillation was found to be independent of the number of atoms for a  fixed number of droplets in the supersolid  but increased with  the number of droplets in the supersolid. For the same atomic interaction parameters ($a$ and $a_{\mathrm{dd}}$) and trap frequencies, the present results for scissors-mode frequency are in excellent agreement with those of Refs. \cite{sci-y,string}.}  The present study of scissors-mode oscillation is complementary to that of Refs. \cite{sci-y,string}, where the authors studied the effect of the variation of the atomic interaction on the scissors-mode oscillation. In this paper we studied the effect of the variation of the trap parameters and the number of atoms and droplets  on the scissors-mode oscillation. { Specifically, we studied the evolution of the scissors-mode oscillation with a variation of the angular frequency $\omega_y$ in the $y$ direction, while the  trap passes from a quasi-1D  to a quasi-2D type.}
  The results of the present study can readily be tested  using the experimental set-up of Ref. \cite{sci-y}.

\begin{acknowledgments}
SKA acknowledges support by the CNPq (Brazil) grant 301324/2019-0.  The use of the supercomputing  cluster of the Universidad de Cartagena is acknowledged.

\end{acknowledgments}



\end{document}